\documentclass[
	aps,
	prd, 
	onecolumn, 
	a4paper, 
	10pt, 
	amsmath,
	amssymb, 
	preprintnumbers,  
	tightenlines, 
	notitlepage,
	floatfix,
	longbibliography
	]{revtex4-1}\pdfoutput=1
\usepackage[utf8]{inputenc}
\usepackage[english]{babel}
\usepackage{bbm}
\usepackage{graphicx}
\usepackage{hyperref}
\usepackage{wasysym}
\usepackage{color}	

\graphicspath{{pics/}}

\allowdisplaybreaks[0]

\input{CQGformat.input}
\input{commands.input}

\newcommand{\eps}{\epsilon}
\newcommand{\osc}[1]{\{ #1\} }					
\newcommand{\gsf}{\mathsf{F}}

\begin{document}

\title{Fast Self-forced Inspirals}

\author{Maarten \surname{van de Meent}}
\ead{mmeent@aei.mpg.de}
\address{Max Planck Institute for Gravitational Physics (Albert Einstein Institute), Potsdam-Golm, Germany}

\author{Niels \surname{Warburton}}
\ead{niels.warburton@ucd.ie}
\address{School of Mathematics and Statistics, University College Dublin, Belfield, Dublin 4, Ireland}

\date{\today}
\begin{abstract}
We present a new, fast method for computing the inspiral trajectory and gravitational waves from extreme mass-ratio inspirals that can incorporate all known and future self-force results. Using near-identity (averaging) transformations we formulate equations of motion that do not explicitly depend upon the orbital phases of the inspiral, making them fast to evaluate, and whose solutions track the evolving constants of motion, orbital phases and waveform phase of a full self-force inspiral with errors of at most order $O(\eta)$, where $\eta$ is the small mass ratio. As a concrete example, we implement these equations for inspirals of non-spinning binaries. Our code computes inspiral trajectories in milliseconds which, depending on the mass-ratio, is a speed up of 2-5 orders of magnitude over previous self-force inspiral models which take minutes to hours to evaluate. Computing two-year duration waveforms using our new model we find a mismatch smaller than $\sim 10^{-4}$ with respect to waveforms computed using slower full self-force models. The speed of our new approach is comparable with kludge models but has the added benefit of easily incorporating self-force results which will, once known, allow the waveform phase to be tracked to sub-radian accuracy over an inspiral. 
\end{abstract}

\maketitle
\setlength{\parindent}{0pt} 
\setlength{\parskip}{6pt}

\section{Introduction}

Deducing the parameters of gravitational-wave sources requires accurate theoretical waveform templates. One challenging class of systems to model are extreme mass-ratio inspirals (EMRIs), a key source for future space-based detectors such as LISA \cite{LISAproposal}. These binary systems, composed of a stellar mass compact object in orbit about a massive $10^5-10^7M_\odot$ black hole, will radiate tens to hundreds of thousands of gravitational wave cycles whilst in the millihertz band of the detector \cite{Babak:2017tow}. These sources, unlike the compact binaries detected with ground-based detectors \cite{Abbott:2016blz,TheLIGOScientific:2017qsa}, will generally not have circularized resulting in a complicated waveform with a rich morphology \cite{Drasco:2005kz}.

For a typical EMRI we expect the gravitational-wave strain induced in the detector to have a very low instantaneous signal-to-noise ratio (SNR). Instead the data can be processed using matched filtering techniques which allow the build-up of the SNR over time. This approach involves comparing the signal against expected theoretical waveform templates. Ideally the waveform templates will have two important properties. First, to avoid significant loss of SNR, they need to be accurate, ideally with a phase accuracy of a fraction of a radian over hundreds of thousands of wave cycles. Second, due to the large parameter space of possible EMRI configurations, they need to be rapid to generate, ideally on a sub-second timescale. 

The `self-force' and `kludge' modeling approaches have arisen to meet these requirements, focusing on either accuracy or speed of computation, respectively. There is some overlap between these two methods and both are based in black hole perturbation theory, which expands the Einstein field equations in powers of the mass ratio around an analytically known black hole solution. 

The primary aim of the self-force approach is to reach the sub-radian accuracy goal. Obtaining this level of accuracy requires calculating the local radiation reaction force, or `self-force' \cite{Hinderer:2008dm}. The equations of motion and regularization procedures employed by this method are now well understood \cite{Poisson:2011nh,Pound:2013faa,Merlin:2016boc,vandeMeent:2017fqk} and many concrete calculations have been made recently \cite{Barack:2010tm,Akcay:2013wfa,Osburn:2014hoa,vandeMeent:2016pee,vandeMeent:2017bcc}. Depending on the orbital configuration, computing the self-force at an instance along a worldline takes minutes to days and even if the self-force along all possible worldlines is precomputed solving the equations of motion can take minutes or hours due to the need to resolve oscillations in the inspiral trajectory on the orbital timescale.

On the other hand, the primary goal of the `kludge' approach is a rapid speed of computation \cite{Barack:2003fp,Babak:2006uv,Chua:2017ujo}. The inspiral is computed by combining fits to (orbit-averaged) numerical flux data with post-Newtonian expansions. As the equations of motion only depend on orbit-averaged quantities there is no need to resolve the inspiral on the orbital timescale. This results in a very rapid computation of the inspiral, with the tradeoff that it does not capture the physics necessary to reach the sub-radian accuracy goal. Initially developed to scope out the data analysis task these models have, over the years, been improving in accuracy by incorporating ever more physics \cite{Huerta:2008gb,Huerta:2011kt}.

In this work we develop and implement a new framework for computing EMRI waveform templates that can both easily incorporate current, and future, self-force results, and also be evaluated on a timescale comparable to kludge models. We achieve this by applying near-identity (averaging) transformations (NITs) to the self-forced equations of motion. Before we describe this technique let us first discuss why self-force inspirals are slow to evaluate.

Within the self-force approach the secondary is treated as a point particle and the inspiral trajectory is computed by calculating the (self-)force this particle experiences due to its interaction with its own metric perturbation. At each instant the self-force is a functional of the past, inspiralling, worldline because radiation that was emitted at an earlier time can backscatter off the spacetime curvature to interact with the particle later on. This dependence on the history of the particle is what makes the self-force challenging to calculate. A self-consistent inspiral can be computed by directly coupling the equations of motion and the field equations and this has been achieved for a toy model involving a scalar charge in orbit about a Schwarzschild black hole \cite{Diener:2011cc} (stability issues have so far prevented a similar calculation in the gravitational case \cite{Dolan:2012jg}). This method is very slow to compute making it infeasible for generating banks of waveform templates (but still important as the gold-standard against which faster methods can be tested). An alternative approach is to approximate the self-force at each instant by the self-force for a particle moving along the unique geodesic tangent to the worldline at that instant \cite{Pound:2007th,Warburton:2011fk,Osburn:2015duj,Warburton:2017sxk}. The advantage of this approach is that the self-force can be computed and interpolated across a, now finite dimensional, parameter space in a preprocessing step. Using the geodesic self-force, rather than the self-force computed using the true inspiral, reduces the equations of motion to a finite dimensional phase space and it is to these equations of motion that we apply our method. The error in the gravitational wave phase induced by using the geodesic, rather then the true, self-force contributes at $\mathcal{O}(\eta^0)$. Formally, this  contributes to the second-order in the mass-ratio dissipative corrections, but initial calculations in the scalar-field case suggest the coefficient of this correction term is small \cite{Warburton_Capra20}.

The crucial feature of the equations of motion that makes numerically finding their solution slow is that they depend explicitly upon the orbital phase(s). As a consequence, a numerical integrator must resolve features of the inspiral on the orbital timescale. With a typical EMRI undergoing on the order of $10^5$-$10^6$ orbits whilst in the detector band, resolving the orbital timescale results in inspiral calculations than take minutes to hours depending upon the mass-ratio of the binary \cite{Osburn:2015duj}. We circumvent this problem by transforming the equations of motion to a new set of variables via a near identity transformation. This transformation has two important properties: i) the resulting equations of motion no longer depend explicitly on the orbital phase and ii) the transformation is small (hence `near identity') such that the solution to the transformed equations of motion remains always close to the solution to the original equations of motion. The first of these properties allows the transformed equations of motion to be numerically solved in milliseconds, rather than minutes or hours as for the original equations. The second property ensures that the resulting solution encapsulates all the self-force physics that the original, slow to compute, solution did. The explicit form of the NIT is derived by positing a general form, with undetermined functions, for a transformation which obeys the second property (that the transformation is `small'), substituting into the original equations of motion, and changing variables before finally choosing the undetermined functions such that they cancel the dependence on the orbital phase in the equations of motion.

Near identity transformations are not new. They have a rich history being applied to dynamical systems, and in particular planetary dynamics, stretching back more than a century~\cite{delaunay1860theorie,vonZeipel1916}. Sometimes called near-identity averaging transformations, the effect of the transformation is to average over the short-timescale physics to produce an equation of motion which captures the long-term secular evolution of the system without the need to resolve the shorter timescale. Averaged equations of motion such as this are ideally suited to the EMRI problem where the main concern is accurately tracking the long-term evolution of the waveform phase. Near identity transformations are closely related to two-timescale expansions, which have been applied in both the PN \cite{Damour:2004bz,Klein:2013qda,Will:2016pgm,Chatziioannou:2016ezg} and self-force regimes \cite{Hinderer:2008dm,Pound:2007ti,Mino:2008rr}. The two methods produce equivalent results, but sometimes one is easier to use. It also seems likely that there is a close relation with the dynamical renormalization group methods of \cite{Galley:2016zee} when applied to an expansion in the mass-ratio.

The existence of an averaging NIT depends only on minimal conditions on the form of the equations of motion which can always be achieved when the unperturbed system is integrable. When an averaging NIT exists it is normally not unique. In Sec.~\ref{sec:general_NIT} we derive a general form for the NITs and discuss different choices that can be made for their form. The NIT method is applicable to inspirals in Kerr spacetime away from orbital resonances, but as a first implementation we compute generic inspirals in Schwarzschild spacetime. Our particular choice of variables, implementation details and waveform generation approach are discussed in Sec.~\ref{sec:schwarzschild}. In Sec.~\ref{sec:results} we discuss our numerical results, showing that the solution to the transformed equations of motion remain close to the solution to the original self-force equation of motion and that the new equations of motion can be solved orders of magnitude more quickly than the originals. We also compute quadrupole waveforms from the NIT and full self-force inspirals and show that the overlap between the two is excellent. We finish with some concluding remarks in Sec.~\ref{sec:conclusion}. Throughout this article we use geometric units such that the speed of light and the gravitational constant are equal to unity.

\section{Averaged equations of motion}\label{sec:general_NIT}
In this section, we derive the near identity transform needed to produce averaged equations of motion for a very generic system. Because of the general nature, this discussion is quite abstract. Readers more interested in the results could skip ahead to the summary in Sec.~\ref{sec:NITsummary}.

\subsection{EMRI equations of motion}
We start from the self-force corrected equations of motion for an extreme mass-ratio inspiral in first-order form,
\begin{subequations}\label{eq:eom}
\begin{align}
\dot{P}_j &= 0 +\eps F^{(1)}_j(\vec{P},\vec{q}) +\eps^2  F^{(2)}_j(\vec{P},\vec{q}) +\bigO(\eps^3),								\\
\dot{q}_i &= \Omega_i(\vec{P}) +\eps f^{(1)}_i(\vec{P},\vec{q}) +\eps^2  f^{(2)}_i(\vec{P},\vec{q}) +\bigO(\eps^3),				\\
\dot{S}_k &= s_k^{(0)}(\vec{P},\vec{q}) +\eps s_k^{(1)}(\vec{P},\vec{q}) +\eps^2  s_k^{(2)}(\vec{P},\vec{q}) +\bigO(\eps^3),
\end{align}
\end{subequations}
where $\eps$ is some small parameter, which we purposefully leave unspecified. The obvious choice would be the small mass-ratio $\mr := m_2/m_1$, but we could also take the symmetric mass-ratio (or something else).

The $\vec{P}=\{P_1,\ldots,P_{j_\mathrm{max}}\}$ is some set of ``geodesic'' constants of motion (i.e., quantities that do not change along a geodesic), which together specify a zeroth-order trajectory in phase space. These could be the actions, energy, angular momentum, eccentricity, angle between the secondary spin and total angular momentum, etc. This set can also include quantities that only acquire evolutionary terms at second order such as the primary mass and spin.

The $\vec{q}=\{q_1,\ldots,q_{i_\mathrm{max}}\}$ are some set of ``phases'' that specify where along a zeroth-order trajectory the system currently is. Together  $\vec{P}$ and $\vec{q}$ should uniquely specify a point in phase space for the system. We require these phase to satisfy two properties: 1)~All functions on the RHS are $2\pi$ periodic in these phases, 2)~the zeroth-order term in their evolution equation (i.e., their ``frequencies'', $\vec{\Omega}$), are independent of the phases $\vec{q}$ themselves. Such a choice is guaranteed to exist if the zeroth-order system is integrable (such as the equations of motion for a test gyroscope in Kerr spacetime), in which case action-angle variables will satisfy the required property \cite{Hinderer:2008dm}. However, we stress that any other choice that satisfies the required properties will work for us.

The $\vec{S}=\{S_1,\ldots,S_{k_\mathrm{max}}\}$ are a set of quantities that are extrinsic to the EMRI's dynamics in the sense that the RHS functions in the evolution equations do not depend on them. They may or may not be extrinsic to the binary itself. The most relevant examples here are the $t$ and $\phi$ coordinates of the secondary. Due to symmetries of the background spacetime they cannot appear explicitly in the equations of motion. Besides these it can also include truly extrinsic quantities such as the center-of-mass velocity of the binary.

Finally, the over dots represent differentiation with respect to some ``time'' parameter used for the evolution of the inspiral. This could be the background Boyer-Lindquist time coordinate, proper time, or something more abstract such as Mino time \cite{Mino:2003yg}.

\subsection{Near identity transform}
Our objective is to rewrite \eqref{eq:eom} in a form where the right hand side is completely independent of the phases  $\vec{q}$. For this we use a tool that has a long history in the study of dynamical systems --- and planetary dynamics in particular --- the \emph{near identity transform} (NIT). This type of transform has previously appeared, though not necessarily by this name, in various studies of EMRIs \cite{vandeMeent:2013sza,vandeMeent:2014raa,Vines:2015efa,Fujita:2016igj}. The presentation here closely follows that of Kevorkian and Cole \cite{KC:1996}. Focusing on the intrinsic quantities first (the extrinsic quantities $\vec{S}$ will be dealt with in section \ref{sec:ext}), the idea is to introduce a small transformation of our phase space coordinates, 
\begin{subequations}\label{eq:nit}
\begin{align}
\tilde{P}_j &= P_j +\eps Y^{(1)}_j(\vec{P},\vec{q}) +\eps^2  Y^{(2)}_j(\vec{P},\vec{q}) +\bigO(\eps^3),					\\
\tilde{q}_i &= q_i +\eps X^{(1)}_i(\vec{P},\vec{q}) +\eps^2  X^{(2)}_i(\vec{P},\vec{q}) +\bigO(\eps^3),
\end{align}
\end{subequations}
where we require the $X^{(n)}_i$ and  $Y^{(n)}_j$ to be smooth periodic functions of the phases $\vec{q}$. Consequently, the difference between the tilded and untilded variables will always be $\bigO(\eps)$, anywhere in the phase space.

The inverse transformation is easily derived by requiring that that the composition with the original transformation is the identity and working order by order in $\eps$,
\begin{subequations}\label{eq:init}
\begin{align}
q_i &= \tilde{q}_i -\eps X^{(1)}_i(\vec{\tilde{P}},\vec{\tilde{q}}) 				\\
&\quad-\eps^2  \Bh{X^{(2)}_i(\vec{\tilde{P}},\vec{\tilde{q}}) -\pd{X^{(1)}_i(\vec{\tilde{P}},\vec{\tilde{q}})}{\tilde{P}_j}Y^{(1)}_j(\vec{\tilde{P}},\vec{\tilde{q}})
-\pd{X^{(1)}_i(\vec{\tilde{P}},\vec{\tilde{q}})}{\tilde{q}_k}X^{(1)}_k(\vec{\tilde{P}},\vec{\tilde{q}})
}
+\bigO(\eps^3), \notag\\
P_j &= \tilde{P}_j -\eps Y^{(1)}_j(\vec{\tilde{P}},\vec{\tilde{q}})					\\
&\quad -\eps^2   \Bh{
Y^{(2)}_j(\vec{\tilde{P}},\vec{\tilde{q}}) -\pd{Y^{(1)}_j(\vec{\tilde{P}},\vec{\tilde{q}})}{\tilde{P}_k}Y^{(1)}_k(\vec{\tilde{P}},\vec{\tilde{q}})
-\pd{Y^{(1)}_j(\vec{\tilde{P}},\vec{\tilde{q}})}{\tilde{q}_k}X^{(1)}_k(\vec{\tilde{P}},\vec{\tilde{q}})
}+\bigO(\eps^3).\notag
\end{align}
\end{subequations}

\subsection{Transformed equations of motion}
By taking the time derivative of the NIT \eqref{eq:nit}, substituting the EMRI equations of motion \eqref{eq:eom} and inverse NIT \eqref{eq:init}, and expanding in powers of $\eps$ we obtain the NIT transformed equations of motions
\begin{subequations}\label{eq:teom}
\begin{align}
\dot{\tilde{P}}_j &= 0 +\eps \tilde{F}^{(1)}_j(\vec{\tilde{P}},\vec{\tilde{q}}) +\eps^2  \tilde{F}^{(2)}_j(\vec{\tilde{P}},\vec{\tilde{q}}) +\bigO(\eps^3),			\\
\dot{\tilde{q}}_i &= \Omega_i(\vec{\tilde{P}}) +\eps \tilde{f}^{(1)}_i(\vec{\tilde{P}},\vec{\tilde{q}}) +\eps^2  \tilde{f}^{(2)}_i(\vec{\tilde{P}},\vec{\tilde{q}}) +\bigO(\eps^3),
\end{align}
\end{subequations}
with

\begin{subequations}\label{eq:ftilde}
\begin{align}
\label{eq:F1tilde}
\tilde{F}^{(1)}_j &=  F^{(1)}_j +\pd{Y^{(1)}_j}{\tilde{q}_i}\Omega_i ,\\
\label{eq:f1tilde}
\tilde{f}^{(1)}_i &=  f^{(1)}_i +\pd{X^{(1)}_i}{\tilde{q}_k}\Omega_k - \pd{\Omega_i}{\tilde{P}_j} Y^{(1)}_j 
,
\end{align}
\end{subequations}
and
\begin{subequations}\label{eq:Ftilde}
\begin{align}
\tilde{F}^{(2)}_j &= 
 	F^{(2)}_j+ \pd{Y^{(2)}_j}{\tilde{q}_i}\Omega_i + \pd{Y^{(1)}_j}{\tilde{q}_i} f^{(1)}_i+ \pd{Y^{(1)}_j}{\tilde{P}_k} F^{(1)}_k 
 	-\pd{\tilde{F}^{(1)}_j}{\tilde{P}_k}Y^{(1)}_k - \pd{\tilde{F}^{(1)}_j}{\tilde{q}_k}X^{(1)}_k, 			\\
\tilde{f}^{(2)}_i &=  f^{(2)}_i+ \pd{X^{(2)}_i}{\tilde{q}_k} \Omega_k+ \pd{X^{(1)}_i}{\tilde{q}_k}  f^{(1)}_k+ \pd{X^{(1)}_i}{\tilde{P}_j}  F^{(1)}_j
  -\pd{\tilde{f}^{(1)}_i}{\tilde{P}_k}Y^{(1)}_k - \pd{\tilde{f}^{(1)}_i}{\tilde{q}_k}X^{(1)}_k				\\
 \notag &\qquad
 -\frac{1}{2} \pdd{\Omega_i}{\tilde{P}_j}{\tilde{P}_k}Y^{(1)}_j Y^{(1)}_k
 - \pd{\Omega_i}{\tilde{P}_j}Y^{(2)}_j
.
\end{align}
\end{subequations}
Here all functions on the right hand side are evaluated at $\vec{\tilde{P}}$ and $\vec{\tilde{q}}$, and we have adopted the convention that all repeated roman indices are summed over.

\subsection{Cancellation of oscillating terms at \texorpdfstring{$\bigO(\eps)$}{O(ϵ)} }
To proceed it is useful to distinguish between oscillating and average pieces of functions. For this we recall that all functions appearing on the RHS of the equations of motion are $2\pi$ periodic in all the phases. Consequently, we can decompose them into Fourier modes. If $A(\vec{P},\vec{q})$ is such a function then we write its Fourier expansion,
\begin{equation}
A(\vec{P},\vec{q}) = \sum_{\vec{\kappa}\in\ZZ^{i_\mathrm{max}}} A_{\vec{\kappa}}(\vec{P})e^{\ii \vec{\kappa}\cdot\vec{q}}.
\end{equation}
Based on this we can define decomposition of $A$ in an average and an oscillatory part
\begin{equation}\label{eq:Fourier_decomp}
A(\vec{P},\vec{q}) = \avg{A}(\vec{P}) + \breve{A}(\vec{P},\vec{q}),
\end{equation}
with
\begin{align}
 \avg{A}(\vec{P}) 			&:=  A_{\vec{0}}(\vec{P}),			\\
\breve{A}(\vec{P},\vec{q})	&:=\sum_{\vec{\kappa}\neq\vec{0}} A_{\vec{\kappa}}(\vec{P})e^{\ii \vec{\kappa}\cdot\vec{q}}.
\end{align}
Using this notation the expression for $F_j^{(1)}$ becomes
\begin{align}
\tilde{F}^{(1)}_j &=  F^{(1)}_j +\pd{Y^{(1)}_j}{\tilde{q}_i}\Omega_i \\
&=  F^{(1)}_j +\pd{\breve{Y}^{(1)}_j}{\tilde{q}_i}\Omega_i\\
&= \avg{F^{(1)}_j} + \sum_{\vec{\kappa}\neq\vec{0}} \hh{F^{(1)}_{j,\vec{\kappa}}+ \ii \bh{\vec{\kappa}\cdot\vec{\Omega}} Y^{(1)}_{j,\vec{\kappa}} }e^{\ii \vec{\kappa}\cdot\vec{q}}.
\end{align}
Consequently, we can eliminate the oscillatory part of $F_j^{(1)}$ by choosing the the oscillatory part of $Y_j^{(1)}$ such that
\begin{equation}\label{eq:Y1sol}
 Y_{j,\vec{\kappa}}^{(1)}(\vec{P}) := \frac{\ii}{\vec{\kappa}\cdot\vec{\Omega}}F_{j,\vec{\kappa}}^{(1)}(\vec{P}).
\end{equation}
Obviously, this choice is only possible if $\vec{\kappa}\cdot\vec{\Omega}\neq 0$ for all $\vec{\kappa}$ such that $F_{j,\vec{\kappa}}^{(1)}\neq 0$. The surfaces in orbital phase space that fail to satisfy this condition are known as orbital resonances. Evolving through these points requires a separate treatment \cite{Flanagan:2010cd,vandeMeent:2013sza}. For this work,  we will assume no resonances occur along the inspiral. This is true generically if the primary black hole has no spin, or for equatorial or spherical inspirals into a spinning black hole. In all three cases the coefficients of the offending terms vanish by the virtue that the forcing terms depend on at most one orbital phase. However, for generic inspirals (featuring both eccentricity and inclination) into a spinning black hole resonances will have to be dealt with.

With the above choice for $\breve{Y}^{(1)}_j$ the expression for $\tilde{f}^{(1)}_i$ becomes
\begin{align}
 \tilde{f}^{(1)}_i
 &=f^{(1)}_i -\pd{\Omega_i}{P_j} Y^{(1)}_j +\pd{X^{(1)}_i}{q_k}\Omega_k \\
 &= \avg{f_i^{(1)}}  -\pd{\Omega_i}{P_j}\avg{Y_j^{(1)}} 
 +\sum_{\vec{\kappa}\neq 0}\hh{ f_{i,\vec{\kappa}}^{(1)}
 - \frac{\ii}{\vec{\kappa}\cdot\vec{\Omega}} \pd{\Omega_i}{P_j} F_{j,\vec{\kappa}}^{(1)}
 + \ii \hh{\vec{\kappa}\cdot\vec{\Omega}} X_{i,\vec{\kappa}}^{(1)}}e^{\ii \vec{\kappa}\cdot\vec{q}}.
\end{align}
Consequently, (in the absence of resonances) we can eliminate the oscillatory part of  $\tilde{f}^{(1)}_i$ by choosing the oscillatory part of  $\tilde{X}^{(1)}_i$ such that
\begin{equation}\label{eq:X1sol}
 X_{i,\vec{\kappa}}^{(1)}(\vec{P}) =
 \frac{\ii}{\vec{\kappa}\cdot\vec{\Omega}}f_{i,\vec{\kappa}}^{(1)}(\vec{P})
 +\pd{\Omega_i}{P_j} \frac{1}{(\vec{\kappa}\cdot\vec{\Omega})^2}F_{j,\vec{\kappa}}^{(1)}(\vec{P}).
\end{equation}

\subsection{Cancellation of oscillating terms at \texorpdfstring{$\bigO(\eps^2)$}{O(ϵ2)} }
With the choice for $\breve{Y}^{(1)}_j$ above the oscillatory part of the expression for $\tilde{F}^{(2)}_j$ becomes,
\begin{align}
\breve{\tilde{F}}^{(2)}_j &= 
 	\breve{F}^{(2)}_j+ \pd{\breve{Y}^{(2)}_j}{\tilde{q}_i}\Omega_i 
 	+ \osc{\pd{\breve{Y}^{(1)}_j}{\tilde{q}_i} f^{(1)}_i}
 	+ \osc{\pd{Y^{(1)}_j}{\tilde{P}_k} F^{(1)}_k}
 	-\pd{\avg{F^{(1)}_j}}{\tilde{P}_k}\breve{Y}^{(1)}_k \\
 	&= \sum_{\vec{\kappa}\neq 0}\Bh{
F^{(2)}_{j,\vec{\kappa}}+ \ii\bh{\vec{\kappa}\cdot\vec{\Omega}}Y^{(2)}_{j,\vec{\kappa}}
	+ \pd{\avg{Y^{(1)}_j}}{\tilde{P}_k} F^{(1)}_{k,\vec{\kappa}}	
 	-\ii\pd{\avg{F^{(1)}_j}}{\tilde{P}_k}\frac{F^{(1)}_{k,\vec{\kappa}}}{\vec{\kappa}\cdot\vec{\Omega}}	\\
\notag
&\qquad 	
 	+\sum_{\vec{\kappa}'\neq 0}\bh{ 
 	\ii \frac{F^{(1)}_{k,\vec{\kappa}-\vec{\kappa}'}}{\vec{\kappa}'\cdot\vec{\Omega}}
 	\Bh{
 		 \pd{F^{(1)}_{j,\vec{\kappa}'}}{\tilde{P}_k} 
 		 -\frac{F^{(1)}_{j\vec{\kappa}'}}{\vec{\kappa}'\cdot\vec{\Omega}}
 		  \pd{\bh{\vec{\kappa}'\cdot\vec{\Omega}}}{\tilde{P}_k}
 	}
 	-
	\frac{\vec{\kappa}'\cdot\vec{f}^{(1)}_{\vec{\kappa}-\vec{\kappa}'}}{\vec{\kappa}'\cdot\vec{\Omega}}F^{(1)}_{j,\vec{\kappa}'}
	}
 	}e^{\ii \vec{\kappa}\cdot\vec{q}},
\end{align}
where we introduced the additional notation $\osc{\cdot}$ to denote the oscillatory part of a product of functions. Consequently, when not at a resonance (i.e., $\vec{\kappa}\cdot\vec{\Omega}\neq 0$), we can eliminate the the oscillatory part by choosing the the oscillatory part of $Y_j^{(2)}$ such that
\begin{align}
Y^{(2)}_{j,\vec{\kappa}}&= 
\frac{\ii}{\bh{\vec{\kappa}\cdot\vec{\Omega}}}\Bh{
F^{(2)}_{j,\vec{\kappa}}
	+ \pd{\avg{Y^{(1)}_j}}{\tilde{P}_k} F^{(1)}_{k,\vec{\kappa}}	
 	-\ii\pd{\avg{F^{(1)}_j}}{\tilde{P}_k}\frac{F^{(1)}_{k,\vec{\kappa}}}{\vec{\kappa}\cdot\vec{\Omega}}	\\
\notag
&\qquad 	
 	+\sum_{\vec{\kappa}'\neq 0}\bh{ 
 	\ii \frac{F^{(1)}_{k,\vec{\kappa}-\vec{\kappa}'}}{\vec{\kappa}'\cdot\vec{\Omega}}
 	\Bh{
 		\pd{F^{(1)}_{j,\vec{\kappa}'}}{\tilde{P}_k} 
 		-\frac{F^{(1)}_{j\vec{\kappa}'}}{\vec{\kappa}'\cdot\vec{\Omega}}
 		\pd{\bh{\vec{\kappa}'\cdot\vec{\Omega}}}{\tilde{P}_k}
 	}
 	-
	\frac{\vec{\kappa}' \cdot\vec{f}^{(1)}_{\vec{\kappa}-\vec{\kappa}'}}{\vec{\kappa}'\cdot\vec{\Omega}}F^{(1)}_{j,\vec{\kappa}'}
	}
}.
\end{align}

We continue in similar fashion with the oscillatory part of $\tilde{f}^{(2)}_i$. With the previous choices this reduces to,
\begin{align}
\breve{\tilde{f}}^{(2)}_i &=  \breve{f}^{(2)}_i
+ \pd{\breve{X}^{(2)}_i}{\tilde{q}_k} \Omega_k
+ \osc{\pd{\breve{X}^{(1)}_i}{\tilde{q}_k}  f^{(1)}_k}
+ \osc{\pd{X^{(1)}_i}{\tilde{P}_j}  F^{(1)}_j}
-\pd{\avg{f^{(1)}_i}}{\tilde{P}_k}\breve{Y}^{(1)}_k 
\\
 \notag &\quad
 +\pd{\Omega_i}{\tilde{P}_j}\pd{\avg{Y^{(1)}_j}}{\tilde{P}_k}\breve{Y}^{(1)}_k 
 - \frac{1}{2} \pdd{\Omega_i}{\tilde{P}_j}{\tilde{P}_k}\osc{\breve{Y}^{(1)}_j \breve{Y}^{(1)}_k}
 - \pd{\Omega_i}{\tilde{P}_j}\breve{Y}^{(2)}_j 
 \\
 &=  \sum_{\vec{\kappa}\neq 0}\Bh{
  	f^{(2)}_{i,\vec{\kappa}}
	+ \ii\bh{\vec{\kappa}\cdot\vec{\Omega}} X^{(2)}_{i,\vec{\kappa}}
	+ \pd{\avg{X^{(1)}_i}}{\tilde{P}_j}  F^{(1)}_{j,\vec{\kappa}} 
	- \pd{\avg{f^{(1)}_i}}{\tilde{P}_k} Y^{(1)}_{k,\vec{\kappa}}
	+\pd{\Omega_i}{\tilde{P}_j}\pd{\avg{Y^{(1)}_j}}{\tilde{P}_k}Y^{(1)}_{k,\vec{\kappa}}
	  \\
 \notag &\qquad
	+ \frac{1}{2} \pdd{\Omega_i}{\tilde{P}_j}{\tilde{P}_k} \avg{Y^{(1)}_j} Y^{(1)}_{k,\vec{\kappa}}
	- \pd{\Omega_i}{\tilde{P}_j}Y^{(2)}_{j,\vec{\kappa}} 
\\ \notag &\qquad
 	 +\sum_{\vec{\kappa}'\neq 0}\bh{
		\ii\bh{\vec{\kappa'}\cdot\vec{f}^{(1)}_{\vec{\kappa}-\vec{\kappa}'}}X^{(1)}_{i,\vec{\kappa}'}  
		+\pd{X^{(1)}_{i,\vec{\kappa}'}}{\tilde{P}_j}  F^{(1)}_{j,\vec{\kappa}-\vec{\kappa}'}
		-	\frac{1}{2} \pdd{\Omega_i}{\tilde{P}_j}{\tilde{P}_k} Y^{(1)}_{j,\vec{\kappa}'} Y^{(1)}_{k,\vec{\kappa}-\vec{\kappa}'}
 }
 }e^{\ii \vec{\kappa}\cdot\vec{q}},
\end{align}
where we have left the $Y^{(n)}_i$, and $X^{(1)}_i$ unexpanded if the explicit choices do not lead to a simplification. Consequently, when $\vec{\kappa}\cdot\vec{\Omega}\neq 0$ we can cancel the oscillatory part by choosing $\breve{X}^{(2)}_i$ such that,
\begin{align}
X^{(2)}_{i,\vec{\kappa}} &= \frac{\ii}{\bh{\vec{\kappa}\cdot\vec{\Omega}}}\Bh{
  	f^{(2)}_{i,\vec{\kappa}}
	+ \pd{\avg{X^{(1)}_i}}{\tilde{P}_j}  F^{(1)}_{j,\vec{\kappa}} 
	- \pd{\avg{f^{(1)}_i}}{\tilde{P}_k} Y^{(1)}_{k,\vec{\kappa}}
	+\pd{\Omega_i}{\tilde{P}_j}\pd{\avg{Y^{(1)}_j}}{\tilde{P}_k}Y^{(1)}_{k,\vec{\kappa}}
	  \\
 \notag &\qquad
	+ \frac{1}{2} \pdd{\Omega_i}{\tilde{P}_j}{\tilde{P}_k} \avg{Y^{(1)}_j} Y^{(1)}_{k,\vec{\kappa}}
	- \pd{\Omega_i}{\tilde{P}_j}Y^{(2)}_{j,\vec{\kappa}} 
\\ \notag &\qquad
 	 +\sum_{\vec{\kappa}'\neq 0}\bh{
		\ii\bh{\vec{\kappa'}\cdot\vec{f}^{(1)}_{\vec{\kappa}-\vec{\kappa}'}}X^{(1)}_{i,\vec{\kappa}'}
		+\pd{X^{(1)}_{i,\vec{\kappa}'}}{\tilde{P}_j}  F^{(1)}_{j,\vec{\kappa}-\vec{\kappa}'}
		-	\frac{1}{2} \pdd{\Omega_i}{\tilde{P}_j}{\tilde{P}_k} Y^{(1)}_{j,\vec{\kappa}'} Y^{(1)}_{k,\vec{\kappa}-\vec{\kappa}'}
 }
 }.
\end{align}
\subsection{Freedom in average pieces of transformation}
With the oscillatory pieces removed by the choices in the previous sections, the remaining average parts of the tilded forcing terms become,
\begin{subequations}\label{eq:avgftilde}
\begin{align}
\label{eq:F1avgtilde}
\tilde{F}^{(1)}_j &=  \avg{F^{(1)}_j} ,			\\
\label{eq:f1avgtilde}
\tilde{f}^{(1)}_i &=  \avg{f^{(1)}_i} -\pd{\Omega_i}{\tilde{P}_j} \avg{Y^{(1)}_j}
,
\end{align}
\end{subequations}
and
\begin{subequations}\label{eq:avgFtilde}
\begin{align}
\tilde{F}^{(2)}_j &= 
 	\avg{F^{(2)}_j}
 	+ \avg{\pd{\breve{Y}^{(1)}_j}{\tilde{q}_i} \breve{f}^{(1)}_i}
 	+ \avg{\pd{\breve{Y}^{(1)}_j}{\tilde{P}_k} \breve{F}^{(1)}_k} 
 	+ \pd{\avg{Y^{(1)}_j}}{\tilde{P}_k} \avg{F^{(1)}_k} 
 	- \pd{\avg{F^{(1)}_j}}{\tilde{P}_k}\avg{Y^{(1)}_k},\\
\tilde{f}^{(2)}_i &=  \avg{f^{(2)}_i}
 - \pd{\Omega_i}{\tilde{P}_j}\avg{Y^{(2)}_j}
 + \avg{\pd{\breve{X}^{(1)}_i}{\tilde{q}_k}  \breve{f}^{(1)}_k}
 + \avg{\pd{\breve{X}^{(1)}_i}{\tilde{P}_j}  \breve{F}^{(1)}_j}
 + \pd{\avg{X^{(1)}_i}}{\tilde{P}_j}  \avg{F^{(1)}_j}
\\
 \notag &\qquad
 -\pd{\avg{f^{(1)}_i}}{\tilde{P}_k} \avg{Y^{(1)}_k}
 +\pd{\Omega_i}{\tilde{P}_j}\pd{\avg{Y^{(1)}_j}}{\tilde{P}_k} \avg{Y^{(1)}_k}
 +\frac{1}{2} \pdd{\Omega_i}{\tilde{P}_j}{\tilde{P}_k} \bh{\avg{Y^{(1)}_j}\avg{Y^{(1)}_k}- \avg{\breve{Y}^{(1)}_j \breve{Y}^{(1)}_k}}
 .
\end{align}
\end{subequations}
We have thus achieved our primary goal: effective equations of motion where the forcing terms do not depend on the phases. Beyond this there is still considerable freedom due to the unconstrained average parts of the near identity transformation $\avg{\vec{X}^{(n)}}$ and $\avg{\vec{Y}^{(n)}}$. Various choices can significantly simplify the equations of motion. We discuss some possibilities in the following subsections.

\subsubsection{No average terms in NIT}
The easiest choice is to simply not include any average terms in the near identity transformation, i.e., set $\avg{\vec{X}^{(n)}}=\avg{\vec{Y}^{(n)}}=0$. Unlike some of the options below this choice is available regardless of the particular details of the original equations of motion. With this choice the NIT'd forcing functions become,
\begin{subequations}\label{eq:Fopt1}
\begin{align}
\tilde{F}^{(1)}_j &=  \avg{F^{(1)}_j} 
,\\
\tilde{f}^{(1)}_i &=  \avg{f^{(1)}_i}
,\\
\tilde{F}^{(2)}_j &= 
 	\avg{F^{(2)}_j}
 	+ \avg{\pd{\breve{Y}^{(1)}_j}{\tilde{q}_i} \breve{f}^{(1)}_i}
 	+ \avg{\pd{\breve{Y}^{(1)}_j}{\tilde{P}_k} \breve{F}^{(1)}_k} 
,\\
\tilde{f}^{(2)}_i &= \avg{f^{(2)}_i}
 + \avg{\pd{\breve{X}^{(1)}_i}{\tilde{q}_k}  \breve{f}^{(1)}_k}
 + \avg{\pd{\breve{X}^{(1)}_i}{\tilde{P}_j}  \breve{F}^{(1)}_j}
 - \frac{1}{2} \pdd{\Omega_i}{\tilde{P}_j}{\tilde{P}_k}  \avg{\breve{Y}^{(1)}_j \breve{Y}^{(1)}_k}
.
\end{align}
\end{subequations}

\subsubsection{Elimination of \texorpdfstring{$\vec{\tilde{f}}^{(2)}$}{f(2)} using \texorpdfstring{$\avg{\vec{{X}}^{(1)}}$}{X(1)}}\label{sec:simplechoice}
The Eqs.~\eqref{eq:Fopt1} are already quite simple, except for the expression for $\vec{\tilde{f}}^{(2)}$. We can improve on this by noting that  $\avg{\vec{X}^{(1)}}$ appears in the forcing functions only through $\vec{\tilde{f}}^{(2)}$. Hence we can eliminate $\vec{\tilde{f}}^{(2)}$ by solving a set of uncoupled first order PDEs,
\begin{align}
 \avg{F^{(1)}_j} \pd{\avg{X^{(1)}_i}}{\tilde{P}_j}  &=
  \pd{\Omega_i}{\tilde{P}_j}\avg{Y^{(2)}_j}
 - \avg{f^{(2)}_i}
 - \avg{\pd{\breve{X}^{(1)}_i}{\tilde{q}_k}  \breve{f}^{(1)}_k}
 - \avg{\pd{\breve{X}^{(1)}_i}{\tilde{P}_j}  \breve{F}^{(1)}_j}
 +\pd{\avg{f^{(1)}_i}}{\tilde{P}_k} \avg{Y^{(1)}_k}
\\
 \notag &\qquad
 -\pd{\Omega_i}{\tilde{P}_j}\pd{\avg{Y^{(1)}_j}}{\tilde{P}_k} \avg{Y^{(1)}_k}
 -\frac{1}{2} \pdd{\Omega_i}{\tilde{P}_j}{\tilde{P}_k} \bh{\avg{Y^{(1)}_j}\avg{Y^{(1)}_k}- \avg{\breve{Y}^{(1)}_j \breve{Y}^{(1)}_k}}
.
\end{align}
Although it may not be possible to provide an explicit solution, it is clear that solutions to these PDEs will exist. Given a numerical realization of the RHS, numerical integration of these equations should be straightforward. Combining this choice with $\avg{\vec{{Y}}^{(1)}}=\avg{\vec{{Y}}^{(2)}}=\avg{\vec{{X}}^{(2)}}=0$, we obtain the fairly simple expressions
\begin{subequations}\label{eq:Fopt2}
\begin{align}
\tilde{F}^{(1)}_j &=  \avg{F^{(1)}_j} 
,\\
\tilde{f}^{(1)}_i &=  \avg{f^{(1)}_i}
,\\
\tilde{F}^{(2)}_j &= 
 	\avg{F^{(2)}_j}
 	+ \avg{\pd{\breve{Y}^{(1)}_j}{\tilde{q}_i} \breve{f}^{(1)}_i}
 	+ \avg{\pd{\breve{Y}^{(1)}_j}{\tilde{P}_k} \breve{F}^{(1)}_k}, 		\\
\tilde{f}^{(2)}_i &=  0.
\end{align}
\end{subequations}
This is the choice we will use in the practical implementation in Sec.~\ref{sec:schwarzschild}. We further note that if one would continue the NIT to higher orders in $\eps$, this choice can be made at arbitrary order to eliminate $\vec{\tilde{f}}^{(n)}$ using the freedom in $\avg{\vec{{X}}^{(n-1)}}$.

\subsubsection{Elimination of post-adiabatic dissipative terms using \texorpdfstring{$\avg{\vec{{Y}}^{(n)}}$}{<Y(n)>}}
In the same spirit as the previous option we can try to eliminate $\vec{\tilde{F}}^{(2)}$ using $\avg{\vec{{Y}}^{(1)}}$. This again requires solving a set of first order PDEs, which are now coupled,
\begin{align}
\avg{F^{(1)}_k} \pd{\avg{Y^{(1)}_j}}{\tilde{P}_k} 
 	-\pd{\avg{F^{(1)}_j}}{\tilde{P}_k}\avg{Y^{(1)}_k} + \avg{\pd{\breve{Y}^{(1)}_j}{\tilde{q}_i} \breve{f}^{(1)}_i}
 	+ \avg{\pd{\breve{Y}^{(1)}_j}{\tilde{P}_k} \breve{F}^{(1)}_k} +\avg{F^{(2)}_j}
 	  &=0.
\end{align}
Solutions to these equations should still exist, at the very least locally/numerically. Moreover, if one would continue the NIT to higher orders in $\eps$, this choice can be made at arbitrary order to eliminate $\vec{\tilde{F}}^{(n)}$ using the freedom in $\avg{\vec{{Y}}^{(n-1)}}$. Together with the option of eliminating all $\vec{\tilde{f}}^{(n)}$ terms with $n\geq 2$ using the freedom in $\avg{\vec{{X}}^{(n-1)}}$, this means that in principle (and provided there are no non-perturbative -- e.g., $\ee^{-\alpha/\eps}$ --  terms) we can find NIT'd equations of motion that are linear in $\eps$,
\begin{subequations}
\begin{align}
\dot{\tilde{P}}_j &= 0 +\eps \tilde{F}^{(1)}_j(\vec{\tilde{P}}),				\\
\dot{\tilde{q}}_i &= \Omega_i(\vec{\tilde{P}}) +\eps \tilde{f}^{(1)}_i(\vec{\tilde{P}}).
\end{align}
\end{subequations}
Note that whilst the equations of motion now appear simpler, unless $\pd{\Omega_i}{\tilde{P}_j}=0$, the solutions for $\avg{{{Y}}^{(1)}_j}$ will appear explicitly in the expression for  ${\tilde{f}}^{(1)}_i$. Furthermore, even if $\pd{\Omega_i}{\tilde{P}_j}\neq0$ the solutions for $\avg{{{Y}}^{(1)}_j}$ will appear explicitly in the expressions for the extrinsic parameters -- see Eq.~\eqref{eq:s_tilde1} in the section on the treatment of the extrinsic parameters.

\subsubsection{Elimination of \texorpdfstring{$\vec{\tilde{f}}^{(n)}$}{f(n)} using \texorpdfstring{$\avg{\vec{{Y}}^{(n)}}$}{Y(n)}}
The expressions for $\tilde{f}^{(n)}_i$ all depend on $\avg{\vec{{Y}}^{(n)}}$ only through a term of the form,
\begin{align}
\pd{\Omega_i}{\tilde{P}_j}\avg{Y^{(n)}_j}.
\end{align}
Consequently, if there exists a left-inverse for the matrix $\pd{\Omega_i}{\tilde{P}_j}$ (i.e., if there exists a matrix $A_k^i$ such that $A_k^i\pd{\Omega_i}{\tilde{P}_j} = \delta_k^j$), we can solve the equation $\vec{\tilde{f}}^{(n)}=0$ for $\avg{\vec{{Y}}^{(n)}}$. This choice eliminates all $\vec{\tilde{f}}^{(n)}$ terms, yielding the following forcing functions (with some abuse of notation we write $\pd{\tilde{P}_j}{\Omega_i}$ for the left-inverse of $\pd{\Omega_i}{\tilde{P}_j}$)
\begin{subequations}\label{eq:Fopt3}
\begin{align}
\tilde{f}^{(n)}_i &=  0,\\
\tilde{F}^{(1)}_j &=  \avg{F^{(1)}_j},	\\
\tilde{F}^{(2)}_j &= 
 	\avg{F^{(2)}_j}
 	+ \avg{\pd{\breve{Y}^{(1)}_j}{\tilde{q}_i} \breve{f}^{(1)}_i}
 	+ \avg{\pd{\breve{Y}^{(1)}_j}{\tilde{P}_k} \breve{F}^{(1)}_k} \\
 	&\qquad\notag
 	+ \avg{F^{(1)}_k} \pd{}{\tilde{P}_k}\Bh{\pd{\tilde{P}_j}{\Omega_i}\avg{f^{(1)}_i}}
 	-\pd{\avg{F^{(1)}_j}}{\tilde{P}_k}\pd{\tilde{P}_k}{\Omega_i}\avg{f^{(1)}_i} 
.
\end{align}
\end{subequations}
However, the existence of a left-inverse of $\pd{\Omega_i}{\tilde{P}_j}$ is not always guaranteed. Section \ref{sec:schwarzschild} shows a trivial way in which this can happen. Namely, if one chooses the ``time'' parameter along the trajectory such that one of the $\Omega_i$ is constant as a function of $\vec{P}$, then the rank  of $\pd{\Omega_i}{\tilde{P}_j}$ is smaller then $i_\mathrm{max}$ and no left-inverse exists. Barring that particularly pathological situation, we normally have less (intrinsic) phases than ``constants of motion'' (i.e., $i_\mathrm{max}<j_\mathrm{max}$), because --- due to symmetries of the background --- some phases conjugate to the actions will be extrinsic to the local dynamics. Consequently, we should generically expect the left-inverse of $\pd{\Omega_i}{\tilde{P}_j}$ to exist.

However, one may still worry that the left-inverse of $\pd{\Omega_i}{\tilde{P}_j}$ may fail to exist on local nodes in the parameter space. One particular reason to worry about this, is the occurrence of isofrequency pairs of orbits -- pairs of physically distinct orbits with the same orbital frequencies $\Omega_r$, $\Omega_\theta$, and $\Omega_\phi$ -- in some regions of orbital parameter space, but not others. On the boundary between two such regions $\pd{\Omega_i}{\tilde{P}_j}$ will become singular.
 In \cite{Warburton:2013yj}, the existence of isofrequency orbits in Kerr spacetime was shown when the frequencies are measured w.r.t.~coordinate time (no such pairings seem to exist for Mino time frequencies \cite{vandeMeent:2013sza}). However, unless there are external perturbations that break axisymmetry, the $\phi$-phase is extrinsic to the local dynamics. So we would only need the matrix $\pd{\Omega_i}{\tilde{P}_j}$ two have rank 2 when restricted to $i\in\{r,\theta\}$ in order for the choice in this subsection to be available. We have not proven so, but this seems likely to be satisfied.

A nice aspect of obtaining the forcing functions in the form \eqref{eq:Fopt3} is that it allows one to directly read off the successive terms in the post-adiabatic (PA) expansion of the inspiral from the $\vec{\tilde{F}}^{(n)}$ terms. To evolve the orbit at adiabatic (0PA, $n=1$) order, we just need the average changes of the constants of motion. At 1PA, in addition, we need the local first order self-force correction and the average changes of the constants of motion at second order. In this way, NIT reproduces the results from the two-timescale expansion of \cite{Hinderer:2008dm}.
 
\subsection{Evolution of extrinsic quantities}\label{sec:ext}
We now turn our attention to the evolution the quantities extrinsic to dynamics, $\vec{S}$. Since, by definition, these quantities do not appear explicitly in the equations of motion we only need their equations of motions up to terms of order $\eps$,
\begin{align}
\dot{S}_k &= s_k^{(0)}(\vec{P},\vec{q}) +\eps s_k^{(1)}(\vec{P},\vec{q}) +\bigO(\eps^2).
\end{align}
By substituting the inverse NIT \eqref{eq:init} and re-expanding in $\eps$ we can write this as an equation involving only the NIT'd variables  $\vec{\tilde{P}}$ and $\vec{\tilde{q}}$,
\begin{align}
\dot{S}_k &= s_k^{(0)} +\eps\bh{
 s_k^{(1)} 
 - \pd{s_k^{(0)}}{\tilde{P}_j} Y^{(1)}_j
 - \pd{s_k^{(0)}}{\tilde{q}_q} X^{(1)}_i
 }+\bigO(\eps^2),
\end{align}
where all functions on the RHS are now understood to be functions of $\vec{\tilde{P}}$ and $\vec{\tilde{q}}$.

We would like to recast these equations in an ``averaged'' form that is independent of the dynamic phases $\vec{\tilde{q}}$. To this end we introduce a new set of transformed extrinsic coordinates $\vec{\tilde{S}}$, defined by the transformation,
\begin{align}\label{eq:Strans}
\tilde{S}_k &= S_k + Z_k^{(0)}(\vec{\tilde{P}},\vec{\tilde{q}}) +\eps Z_k^{(1)}(\vec{\tilde{P}},\vec{\tilde{q}}) +\bigO(\eps^2).
\end{align}
Note that this is not a near-identity transform due to the inclusion of the $Z_k^{(0)}$ term at zeroth order. This means that for the production of waveforms it will be necessary to know the details of this transformation.

By taking the time derivative of \eqref{eq:Strans} and substituting the equations of motion for $\vec{S}$ we obtain equations of motion for $\vec{\tilde{S}}$,
\begin{align}
\dot{\tilde{S}}_k = \tilde{s}_k^{(0)} +\eps \tilde{s}_k^{(1)} +\bigO(\eps^2),
\end{align}
with
\begin{align}
\tilde{s}_k^{(0)} &:=  s_k^{(0)} + \pd{Z_k^{(0)}}{\tilde{q}_i}\Omega_i 
,\\
\tilde{s}_k^{(1)} &:=  s_k^{(1)} 
 - \pd{s_k^{(0)}}{\tilde{P}_j} Y^{(1)}_j
 - \pd{s_k^{(0)}}{\tilde{q}_i} X^{(1)}_i
 + \pd{Z_k^{(0)}}{\tilde{q}_i}\tilde{f}^{(1)}_i 
 + \pd{Z_k^{(0)}}{\tilde{P}_j}\tilde{F}^{(1)}_j 
 + \pd{Z_k^{(1)}}{\tilde{q}_i}\Omega_i.
\end{align}

We can eliminate the oscillatory parts of the forcing functions $\tilde{s}_k^{(n)}$ by solving the equations
\begin{align}
\label{eq:Z0eq}
 \breve{s}_k^{(0)} + \pd{\breve{Z}_k^{(0)}}{\tilde{q}_i}\Omega_i &=0
 ,\\
 \breve{s}_k^{(1)} 
 - \osc{\pd{s_k^{(0)}}{\tilde{P}_j} Y^{(1)}_j}
 - \osc{\pd{\breve{s}_k^{(0)}}{\tilde{q}_i} X^{(1)}_i}
 + \pd{\breve{Z}_k^{(0)}}{\tilde{q}_i}\tilde{f}^{(1)}_i 
 + \pd{\breve{Z}_k^{(0)}}{\tilde{P}_j}\tilde{F}^{(1)}_j 
 + \pd{\breve{Z}_k^{(1)}}{\tilde{q}_i}\Omega_i &=0
\end{align}
for the oscillatory parts of the transformation, $\breve{Z}_k^{(n)}$. Solutions for both equations clearly exist. Solving the first equation is akin to solving equations of motion at the test body level, which in many cases can be done analytically. The second equation would have to be solved numerically. However, in practice it is sufficient to know that it exists, since $\breve{s}_k^{(1)}$ will only explicitly appear in the second order forcing term for  $\tilde{S}^{(1)}_k$. Consequently, it will only enter the waveform at order $\bigO(\eps)$, and can thus be neglected.

The remaining forcing functions depend only on $\vec{\tilde{P}}$ and are given by,
\begin{align}
\tilde{s}_k^{(0)} &=  \avg{s_k^{(0)}}
,\\
\tilde{s}_k^{(1)} &=  \avg{s_k^{(1)}} \label{eq:s_tilde1}
 - \pd{\avg{s_k^{(0)}}}{\tilde{P}_j} \avg{Y^{(1)}_j}
 - \avg{\pd{\breve{s}_k^{(0)}}{\tilde{P}_j} \breve{Y}^{(1)}_j}
 - \avg{\pd{\breve{s}_k^{(0)}}{\tilde{q}_i} \breve{X}^{(1)}_i}
 + \pd{\avg{Z_k^{(0)}}}{\tilde{P}_j}\tilde{F}^{(1)}_j.
\end{align}
In principle, it is possible to eliminate the first order forcing term $\tilde{s}_k^{(1)}$ completely by solving a first order linear partial differential equation for $\avg{Z_k^{(0)}}$. However, since $\avg{Z_k^{(0)}}$ will appear explicitly in any construction of the waveform, this is of little utility. Instead it is much easier to just set  $\avg{Z_k^{(0)}}=0$.

\subsection{Summary of NIT results}\label{sec:NITsummary}
Using a set of averaging transformation we have recast the the small mass-ratio expanded equations of motion for a compact binary \eqref{eq:eom} in an orbit averaged form that is independent of the phases,
\begin{subequations}\label{eq:eomavg}
\begin{align}
\dot{\tilde{P}}_j &= 0 +\eps \tilde{F}^{(1)}_j(\vec{\tilde{P}}) +\eps^2  \tilde{F}^{(2)}_j(\vec{\tilde{P}}) +\bigO(\eps^3),\\
\dot{\tilde{q}}_i &= \Omega_i(\vec{\tilde{P}}) +\eps \tilde{f}^{(1)}_i(\vec{\tilde{P}}) +\eps^2  \tilde{f}^{(2)}_i(\vec{\tilde{P}}) +\bigO(\eps^3),\\
\dot{\tilde{S}}_k &= \tilde{s}_k^{(0)}(\vec{\tilde{P}}) +\eps \tilde{s}_k^{(1)}(\vec{\tilde{P}}) + \bigO(\eps^2),
\end{align}
\end{subequations}
The forcing functions are given by
\begin{align}
\tilde{F}^{(1)}_j &=  \avg{F^{(1)}_j}, 
&
\tilde{F}^{(2)}_j &= 
 	\avg{F^{(2)}_j}
 	+ \avg{\pd{\breve{Y}^{(1)}_j}{\tilde{q}_i} \breve{f}^{(1)}_i}
 	+ \avg{\pd{\breve{Y}^{(1)}_j}{\tilde{P}_k} \breve{F}^{(1)}_k}, 
\\
\tilde{f}^{(1)}_i &=  \avg{f^{(1)}_i},
&
\tilde{f}^{(2)}_i &=  0,
\\
\tilde{s}_k^{(0)} &=  \avg{s_k^{(0)}},
&
\tilde{s}_k^{(1)} &=  \avg{s_k^{(1)}} 
 - \avg{\pd{\breve{s}_k^{(0)}}{\tilde{P}_j} \breve{Y}^{(1)}_j}
 - \avg{\pd{\breve{s}_k^{(0)}}{\tilde{q}_i} \breve{X}^{(1)}_i},
\end{align}
where
\begin{align}
 \breve{Y}_{j}^{(1)} &= \sum_{\vec{\kappa}\neq 0} \frac{\ii}{\vec{\kappa}\cdot\vec{\Omega}}F_{j,\vec{\kappa}}^{(1)} \ee^{\ii \vec{\kappa}\cdot\vec{q}}
,\\
\breve{X}_{i}^{(1)} &=
 \sum_{\vec{\kappa}\neq 0} \Bh{\frac{\ii}{\vec{\kappa}\cdot\vec{\Omega}}f_{i,\vec{\kappa}}^{(1)}
 +\frac{1}{(\vec{\kappa}\cdot\vec{\Omega})^2}\pd{\Omega_i}{P_j} F_{j,\vec{\kappa}}^{(1)})}\ee^{\ii \vec{\kappa}\cdot\vec{q}}.
\end{align}
To recover the original variables $(\vec{P}, \vec{q}, \vec{S})$ --- which are needed to construct the generated waveform --- we need to apply the inverse transformation at leading order
\begin{align}
P_j &= \tilde{P}_j +\bigO(\eps),\\
q_i &= \tilde{q}_i +\bigO(\eps),\\
S_k &= \tilde{S}_k -  Z_k^{(0)}(\vec{\tilde{P}},\vec{\tilde{q}}) +\bigO(\eps),
\end{align}
where  $Z_k^{(0)}$ is found by solving \eqref{eq:Z0eq}, preferably analytically.

This analysis independently confirms an important result from the two-timescale analysis of the same system \cite{Hinderer:2008dm}: to evolve the dynamics of the system over an $\bigO(\eps^{-1})$ time making an error in the phases of at most $\bigO(\eps)$, one needs the first order corrections to the equations of motion and the average dissipative corrections at second order.

Finally, we stress an important caveat: the transformation above is only possible if the terms $\frac{F_{j,\vec{\kappa}}^{(1)}}{(\vec{\kappa}\cdot\vec{\Omega})^2}$ and $\frac{f_{i,\vec{\kappa}}^{(1)}}{\vec{\kappa}\cdot\vec{\Omega}}$ stay bounded everywhere along the inspiral. In other words, this procedure works only in the absence of orbital resonances. If resonances do occur, a different analysis is needed in the vicinity of the resonance \cite{Flanagan:2010cd,vandeMeent:2013sza}.

\section{Schwarzschild case}\label{sec:schwarzschild}
The previous section was purposefully very abstract so that it is applicable to, e.g., generic inspirals into a rotating black hole away from orbital resonances. In this section we apply NITs to a concrete evolution problem: the evolution of a non-spinning extreme mass-ratio inspiral under the gravitational self-force.

\subsection{Equations of motion}
Our first task will be to find a set of equations of motion in the form of \eqref{eq:eom}. For this we employ the method of osculating geodesics \cite{Pound:2007th}. At each point in time, the trajectory of the secondary is described by a tangent geodesic in the background Schwarzschild spacetime generated by the primary. To describe such a geodesic we need two constants of motion and one phase. For the two constants of motion we use the semi-latus rectum $p$ and eccentricity $e$. These are defined following Darwin \cite{Darwin:1959,Darwin:1961} using the periapsis and apoapsis distance $r_{\min}$ and $r_{\max}$,
\begin{align}
p &:= \frac{2 \rmin\rmax}{(\rmin+\rmax)M}, & && e &:= \frac{\rmax-\rmin}{\rmax+\rmin}.
\end{align}
where $M$ is the mass of the massive black hole. As the phase we use the relativistic anomaly $\xi$ (also introduced by Darwin \cite{Darwin:1959,Darwin:1961}) defined by the relation
\begin{equation}\label{eq:Darwin_r}
 r = \frac{p M}{1+e\cos\xi}.
\end{equation}
To fully describe the trajectory of the secondary we also need two quantities extrinsic to the dynamics; the coordinate values of $t$ and $\phi$. The osculating geodesics evolution equations in these coordinates were provided by Pound and Poisson \cite{Pound:2007th} and take the form
\begin{subequations}\label{eq:Seom}
\begin{align}
\d{\xi}{\chi} &= 1 + \eta f_\xi(p,e,\xi), 							\\
\d{p}{\chi} &= \eta \mathcal{F}_p(p,e,\xi),							\\
\d{e}{\chi} &= \eta \mathcal{F}_e(p,e,\xi),							\\
\d{t}{\chi} &= \omega_t(p,e,\xi), 				\label{eq:Seom_t}	\\
\d{\phi}{\chi} &= \omega_\phi(p,e,\xi)			\label{eq:Seom_phi},
\end{align}
\end{subequations}
where $\eta$ is the mass ratio $m_2/m_1$ and the ``time'' parameter along the trajectory, $\chi$, is defined such that when $\eta=0$, $d\xi/d\chi =1$. The full details of the functions $f_\xi$, $\mathcal{F}_{p/e}$, and $\omega_{t,\phi}$ are given in \ref{app:pars}. The Eqs.~\eqref{eq:Seom} are of the form \eqref{eq:eom} with $\vec{q} = \{\xi\}$, $\vec{P} = \{p,e\}$, $\vec{S} = \{t,\phi\}$, and $\eps=\eta$. We can thus follow the procedure of Sec.~\ref{sec:general_NIT} (using the choices of Sec.~\ref{sec:simplechoice}) to a find an averaged version of the equations of motion,
\begin{subequations}\label{eq:Seomavg}
\begin{align}
\d{\tilde{\xi}}{\chi} &= 1 
+\eta \tilde{f}_\xi^{(1)}(\tilde{p},\tilde{e})
+\bigO(\eta^3),\\
\d{\tilde{p}}{\chi} &=  
\eta \tilde{\mathcal{F}}_p^{(1)}(\tilde{p},\tilde{e})
+\eta^2 \tilde{\mathcal{F}}_p^{(2)}(\tilde{p},\tilde{e})
+\bigO(\eta^3),
\\
\d{\tilde{e}}{\chi} &= 
\eta \tilde{\mathcal{F}}_e^{(1)}(\tilde{p},\tilde{e}) 
+\eta^2 \tilde{\mathcal{F}}_e^{(2)}(\tilde{p},\tilde{e})
+\bigO(\eta^3),
\\
\d{\tilde{t}}{\chi} &= 
\frac{T_r(\tilde{p},\tilde{e})}{2\pi}
 +\eta \tilde{f}_t^{(1)}(\tilde{p},\tilde{e})
 +\bigO(\eta^2),
 \\
\d{\tilde{\phi}}{\chi} &=
\frac{\Phi_r(\tilde{p},\tilde{e})}{2\pi}
 +\eta \tilde{f}_\phi^{(1)}(\tilde{p},\tilde{e})
 +\bigO(\eta^2),
\end{align}
\end{subequations}
where $T_r(p,e)$ and $\Phi_r(p,e)$ are the radial period and total accumulated $\phi$ over such a period of a Schwarzschild geodesic described by $(p,e)$, and the averaged forcing functions are given by
\begin{subequations}\label{eq:tilde_Fs}
\begin{align}
\tilde{f}_\xi^{(1)} &= \avg{f_\xi},
\qquad\qquad
\tilde{\mathcal{F}}_p^{(1)} =  \avg{ \mathcal{F}_p},
\qquad\qquad
\tilde{\mathcal{F}}_e^{(1)} = \avg{ \mathcal{F}_e},
\\
\tilde{\mathcal{F}}_p^{(2)} &=
 -\avg{\breve{\mathcal{F}}_p \int \pd{ \breve{\mathcal{F}_p}}{p}\id\xi }
 -\avg{\breve{\mathcal{F}}_e \int \pd{\breve{\mathcal{F}_p}}{e}\id\xi }
 -\avg{\breve{\mathcal{F}}_p \breve{f}_\xi},
\\
\tilde{\mathcal{F}}_e^{(2)} &=
 -\avg{\breve{\mathcal{F}}_p\int \pd{\breve{\mathcal{F}_e}}{p} \id\xi }
 -\avg{\breve{\mathcal{F}}_e\int \pd{\breve{\mathcal{F}_e}}{e} \id\xi }
 -\avg{\breve{\mathcal{F}}_e \breve{f}_\xi}, 
\\
\tilde{f}_t^{(1)} &= 
 \avg{\pd{\breve{Z}_t^{(0)}}{p} \breve{F}_p}
+ \avg{\pd{\breve{Z}_t^{(0)}}{e} \breve{F}_e},
\qquad
\tilde{f}_\phi^{(1)} = 
 \avg{\pd{\breve{Z}_\phi^{(0)}}{p} \breve{F}_p}
+ \avg{\pd{\breve{Z}_\phi^{(0)}}{e} \breve{F}_e}.
\end{align}
\end{subequations}
The full details of the NIT need to achieve this form are given in \ref{app:NIT}, where we also give Eq.~\eqref{eq:tilde_Fs} written in terms of Fourier coefficients of the original forcing functions, a form which is particularly useful for practical implementation. In that appendix we also give the analytic formula for $T_r$ and $\Phi_r$.

\subsection{Implementation}

Constructing the functions, $\tilde{\mathcal{F}}^{(1/2)}_{p/e}, \tilde{f}^{(1)}_{\xi/t/\phi}$ on the right-hand side of the NIT equations of motion \eqref{eq:Seomavg} requires knowledge of both the self-force and its derivatives with respect to the $(p,e)$ orbital elements. At present there are no self-force codes that directly compute these derivatives. Instead, we employ an analytic model for the self-force with numerically fitted coefficients in the range $0\le e \le 0.2$ and $6+2e < p \le 12$ from Ref.~\cite{Warburton:2011fk}. The analytic nature of the model makes it straightforward to take derivatives of the self-force with respect to $p$ and $e$. A number of pre-processing, or offline, steps are applied to the full self-force model to construct the NIT inspiral model. These offline steps only need to be computed once. The inspiral trajectory can then be rapidly evaluated in an online step for a given mass-ratio and initial parameters $(p_0, e_0)$. The steps that can be precomputed offline are:

\begin{enumerate}
	
		\item{[Offline] Compute the gravitational self-force along geodesic orbits at many thousands of points in the $(p,e)$ parameter space using codes such as those presented in \cite{Barack:2010tm,Akcay:2013wfa,Osburn:2014hoa}. This step can takes days running on hundreds of processors and produces gigabytes of data. Once all the data is in hand it can be interpolated using a global \cite{Warburton:2011fk} or local \cite{Osburn:2015duj} fit to produce the rapidly evaluated functions $\mathcal{F}_{p/e}$ and $f_{\xi}$. }
		\item{[Offline] Compute the coefficients in the Fourier expansion \eqref{eq:Fourier_decomp} of the functions $\mathcal{F}_{p/e}, f_{\xi/t/\phi}$ on a grid of points in the $(p,e)$ parameter space. We choose to use a grid with regular spacing in $p$ and $e$ as it simplifies the construction of the two-dimensional interpolatants in step (iii). The decomposition into Fourier modes is performed using the efficient FFTW C-library \cite{FFTW05}. With a spacing of $\Delta p = 0.05$ and $\Delta e = 0.002$ this step takes $\sim2.5$ minutes. This step is the first step in the calculation presented in this work as the prior step was carried out in \cite{Warburton:2011fk} and the fit made publicly available.}
		\item{[Offline] Compute the averaged forcing functions $\tilde{\mathcal{F}}^{(1/2)}_{p/e}, \tilde{f}^{(1)}_{\xi/t/\phi}$ at each point in the $(p,e)$ parameter space using the Fourier form of Eq.~\eqref{eq:tilde_Fs} given in Eq.~\eqref{eq:tilde_Fsexplicit}, and save the output to disk. This step takes less than 2 seconds and the stored data takes up $\sim 2$ megabytes of disk space. }
		\item{[Offline] Interpolate the grid of data for each of $\tilde{\mathcal{F}}^{(1/2)}_{p/e}, \tilde{f}^{(1)}_{\xi/t/\phi}$. In our implementation we use cubic spline interpolation from the GNU Scientific Library \cite{GSL}. This step takes $\sim 35$ milliseconds.}
\end{enumerate}
All the times quoted above are computed on a single core of a 2.5GHz MacBook laptop. The online steps that can be computed rapidly for each set of initial conditions are:
\begin{enumerate}
		\setcounter{enumi}{3}
		\item{[Online] Compute an inspiral using Eqs.~\eqref{eq:Seomavg}. In our implementation we solve the ODEs using an adaptive Runge-Kutta algorithm from the GNU Scientific Library \cite{GSL}.}
		\item{[Online] With the inspiral trajectory in hand, the waveform can be computed as outlined in the next subsection.}
\end{enumerate}

We discuss in the results section below the computation time of the online steps. An implementation of the above steps in the C\texttt{++} programming language is publicly available as part of the the Black Hole Perturbation Toolkit \cite{BlackHolePerturbationToolkit}.  The code is licensed under the open-source GNU Public Licence (GPL).

\subsection{Waveform generation}\label{sec:waveform_generation}

In most approaches the method for computing the waveform is independent of the method used to compute the inspiral trajectory. Given an inspiral trajectory there are a number of ways to construct the associated gravitational waveform. The most robust method, but also the most computational expensive, is to use the trajectory as a source in a time-domain perturbation code, such as \cite{Barack:2010tm} (Schwarzschild) or \cite{Sundararajan:2007jg,Harms:2013ib} (Kerr). Computing tens to hundreds of thousands of waveform cycles using this method is infeasible, but for a smaller number of cycles this approach is an important benchmark for the methods outlined below.

One alternative method is to stitch together a sequence of so-called `snapshot' waveforms. Each snapshot is the waveform associated with a particle moving along a bound geodesic. The periodic nature of bound geodesics means these snapshots can be rapidly computed using frequency-domain perturbation codes. These snapshots can be precomputed and interpolated across the parameter space in an offline step. The waveform for a given inspiral can then be constructed by smoothly moving from one snapshot to the next. This method has been implemented in e.g., \cite{Osburn:2015duj}.

Another commonly used waveform generation algorithm is the `semi-relativistic approximation' \cite{Ruffini:1981af} often used by kludge methods \cite{Barack:2003fp,Babak:2006uv,Chua:2017ujo}. In this approach the Schwarzschild (or Boyer-Lindquist) coordinates of the inspiral trajectory are mapped to flat-space coordinates. The waveform is then constructed using the quadrupole formula (possibly with octupolar corrections). Despite the black hole to flat space coordinate map this method has been shown to produce surprisingly accurate results in the strong-field when compared to snapshot waveforms \cite{Babak:2006uv}.

For our purposes it does not matter which waveform generation scheme we use so long as we use the same method with the full self-force and NIT inspiral to allow for a fair comparison. Thus, we opt to use the semi-relativistic approximation in this work as it is the simplest to implement. Details of this method can be found in, e.g., \cite{Gair:2005is,Babak:2006uv}.

\section{Results}\label{sec:results}

The key feature of a NIT inspiral is that it can be evaluated rapidly and at the same time the evolving constants of motion, phases and extrinsic parameters remain within $\mathcal{O}(\eta)$ of an inspiral computed using the full self-force. In this section we present numerical results which demonstrate these two properties of the NIT inspiral. We also show that the waveform computed using the NIT inspiral trajectory is an excellent match with respect to the waveform computed using the full inspiral trajectory.

\begin{figure}
	\includegraphics[width=12cm]{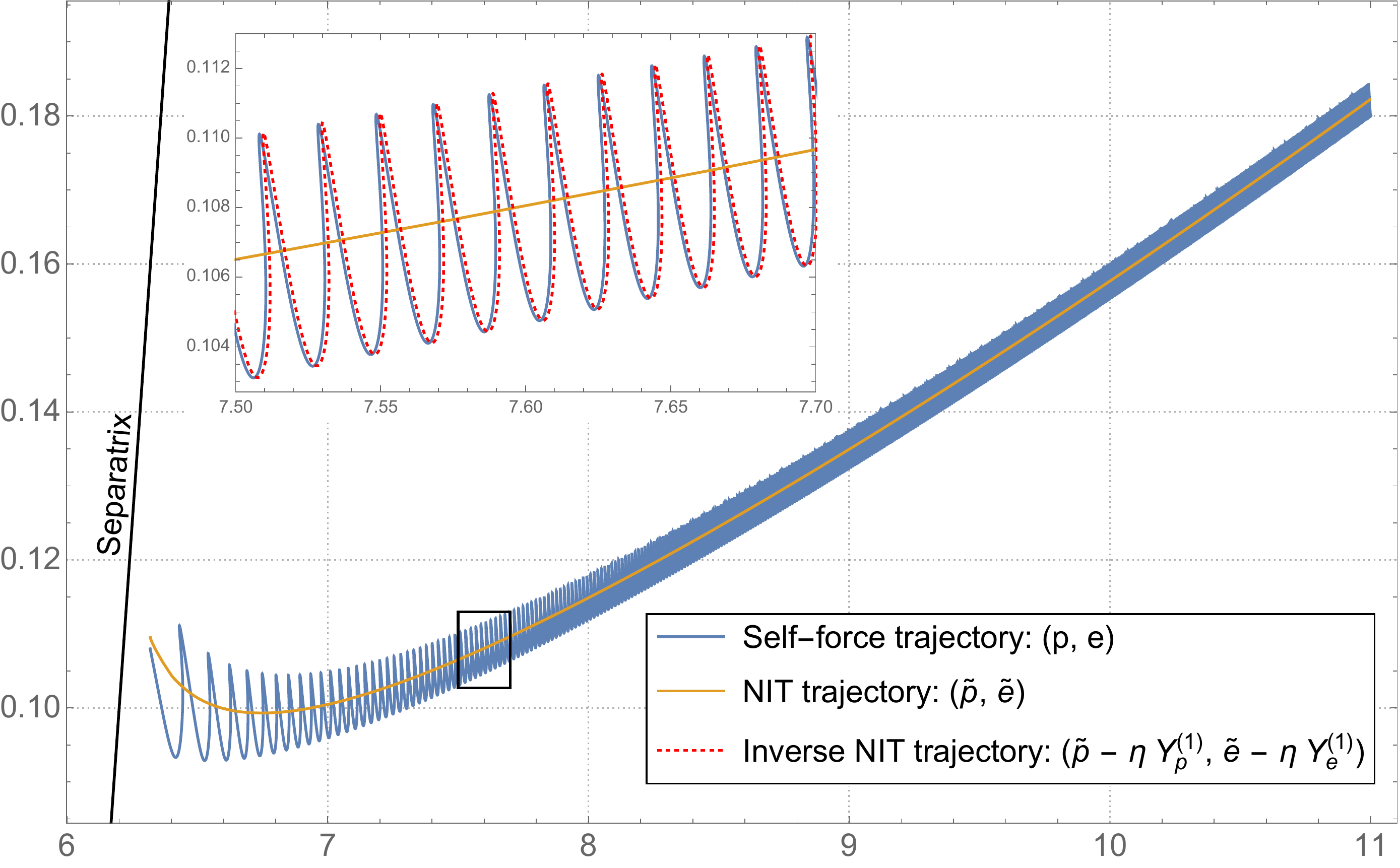}
	\caption{Inspiral trajectory for a binary with $\eta=10^{-3}$ and initial parameters $(p_0,e_0) = (11,0.18)$. This relatively large mass ratio was chosen to make the oscillations in the self-forced inspiral clear. The oscillating (blue) curve shows the trajectory of $(p,e)$ for the self-force inspiral. The smooth (orange) curve shows the trajectory of $(\tilde{p},\tilde{e})$ for the NIT inspiral. The solid (black) line shows the location of the separatrix between bound and plunging orbits. The inset figure shows a zoom in of the region inside the black rectangle. In the inset the dotted (red) curve shows the result of applying the inverse NIT, Eq.~\eqref{eq:init}, through $\mathcal{O}(\eta)$ to the NIT trajectory. The inverse NIT trajectory and the self-force trajectory are in good agreement at this late stage of the inspiral. This agreement improves further for smaller mass ratios.}\label{fig:insp_trajectory}
\end{figure}

First, let us demonstrate the accuracy of the NIT inspiral method. Figure \ref{fig:insp_trajectory} gives an example of the evolution of $(p,e)$ and $(\tilde{p},\tilde{e})$ for the full self-force and NIT inspirals, respectively. We compute the full inspiral trajectory using an osculating element prescription \cite{Pound:2007th} coupled to an interpolated self-force model \cite{Warburton:2011fk}.  The resulting full self-force inspiral trajectory clearly shows oscillations on the orbital timescale which, as discussed in the introduction, is what slows down the numerical computation of the trajectory. To compute the corresponding NIT inspiral we first transform the initial conditions $(p_0,e_0)$ using the first-order NIT Eq.~\eqref{eq:Snit} up to $\bigO(\eta)$ to get $(\tilde{p}_0,\tilde{e}_0)$. We then numerically solve for the NIT inspiral using Eqs.~\eqref{eq:Seomavg}. The NIT inspiral trajectory is then a smooth curve with no oscillations that runs through the `average' of the oscillating full self-force inspiral trajectory. The accuracy of the NIT inspiral trajectory can be illustrated by applying the inverse NIT transformation, Eq.~\eqref{eq:init}, through $\bigO(\eta)$ and comparing to the full self-force trajectory. The inset of Fig.~\ref{fig:insp_trajectory} shows close agreement between the two inspiral trajectories. This comparison improves as the mass ratio is made smaller (we used a relatively large mass-ratio of $\eta=10^{-3}$ for Fig.~\ref{fig:insp_trajectory} to make the oscillations in the full self-force inspiral clear).

\begin{figure}
	\includegraphics[width=12cm]{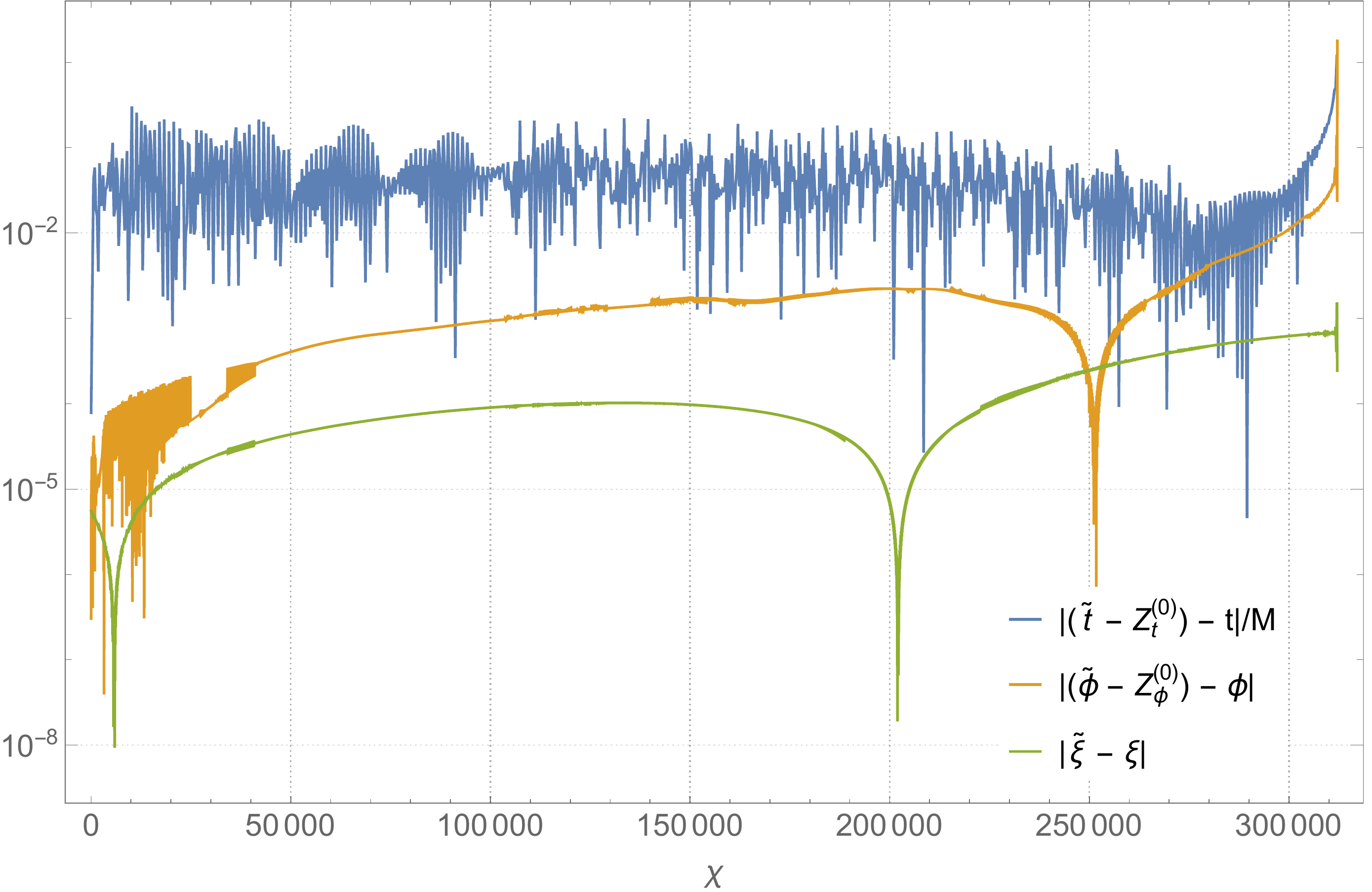}
	\caption{Difference in the full self-force extrinsic parameters $t,\phi$ and phase $\xi$ and their NIT equivalents for a binary with initial parameters $(p_0,e_0) = (11,0.18)$ and mass ratio $\eta = 10^{-5}$. All the variables oscillate on the orbital timescale which is the origin of the noisy features in the curves. Apart from close to plunge, where the NIT breaks down, we find the full self-force and NIT inspiral are in excellent agreement with $|\tilde{\xi}-\xi| \lesssim10^{-3}$, $|(\tilde{t}-Z_t^{(0)}) - t|/M \lesssim 0.1$ and $|(\tilde{\varphi} - Z_\phi^{(0)}) - \varphi| \lesssim10^{-2}$. }\label{fig:phase_difference}
\end{figure}

The evolution of the phase and extrinsic parameters $\{\xi,t,\varphi\}$ show similarly excellent agreement between the NIT and full self-force inspirals. Figure \ref{fig:phase_difference} shows sample results for an inspiral with $\eta=10^{-5}$. We find the difference in the phase, $|\tilde{\xi}-\xi|$, remains less than $10^{-3}$ over the entire inspiral excluding the last few orbits where, with the onset of the plunge, the adiabatic approximation breaks down and with it the effectiveness of the NIT. To compare NIT'd extrinsic parameters, $\{\tilde{t},\tilde{\varphi}\}$, with $\{t,\varphi\}$ one must first restore the $\mathcal{O}(\eta^0)$ oscillatory terms, $Z_t^{(0)}$ and $Z_\phi^{(0)}$. These terms are given analytically in terms of elliptic integrals in Eqs.~\eqref{eq:Z_t0} and \eqref{eq:Z_phi0} and they are quick to evaluate. We find $|(\tilde{t}-Z_t^{(0)}) - t|/M \lesssim 0.1$ over the inspiral up to a few cycles before plunge. For the azimuthal phase we find $|(\tilde{\varphi} - Z_\phi^{(0)}) - \varphi| \lesssim10^{-2}$ radians over most of the inspiral.

\begin{figure}
	\includegraphics[width=15cm]{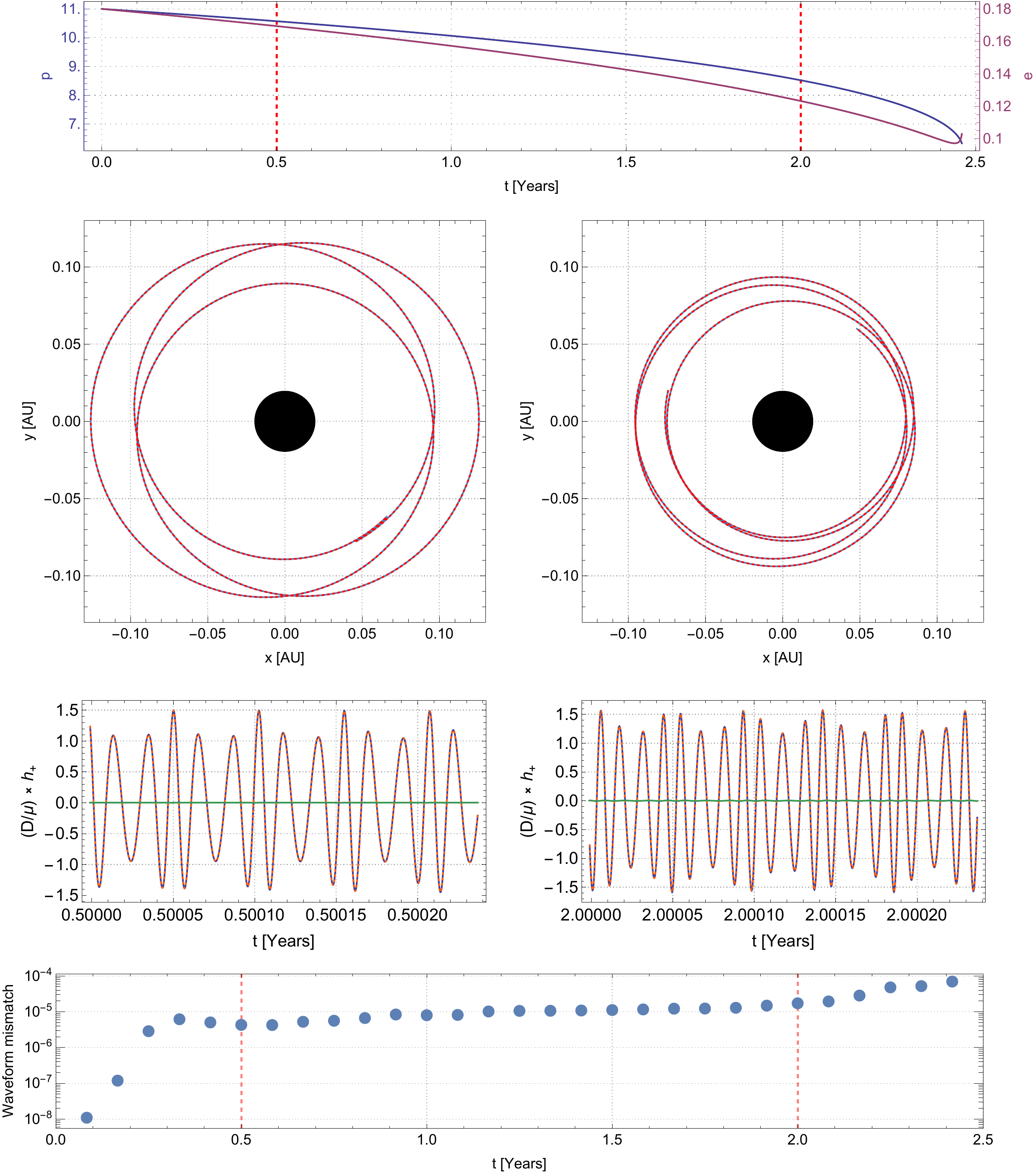}
	\caption{Comparison of a full self-force and NIT inspiral for a binary with $M=10^6 M_\odot$ and $\eta=10^{-5}$. The initial parameters of the full self-force inspiral are $(p_0,e_0) = (11,0.18)$ and the NIT inspiral's initial parameters $(\tilde{p}_0,\tilde{e}_0)$ are computed using the inverse NIT transformation Eq.~\eqref{eq:init} through $\mathcal{O}(\eta)$. The inspiral lasts just under 2.5 years. (Top) the evolution of $p$ (scale on left axis) and $e$ (scale on right axis). (Top middle) Comparison of inspiral trajectories over two orbits after (left) 6 months and (right) 2 years. The (blue) solid curve shows the full self-force trajectory and the (red) dotted curve shows the NIT trajectory. In both plots the difference between the two trajectories is invisible to the eye. (Bottom middle) Comparison of quadrupolar waveforms after (left) 6 months and (right) 2 years. The (blue) solid curve shows the waveform from the full self-force inspiral and the (red) dotted curve shows the waveform from the NIT inspiral. The small amplitude solid (green) curve shows the small difference between the two waveforms. (Bottom) Waveform mismatch computed as a function of time. The mismatch remains small across the entire lifetime of the inspiral.}\label{fig:insp_and_waveform}
\end{figure}

The close agreement between the description of the full self-force inspiral $\{p,e,t,\varphi\}$ and the description of the NIT inspiral $\{\tilde{p}, \tilde{e}, \tilde{t} - Z_t^{(0)}, \tilde{\varphi} - Z_\phi^{(0)}\}$ implies the NIT waveform is a good approximation to the full self-force waveform. To quantify this we compute waveforms using the kludge quadrupole approximation described in Sec.~\ref{sec:waveform_generation} and calculate the mismatch  between the two waveforms, minimizing over phase and time shifts. To do this we use Eq.~(4) of \cite{Ohme:2011zm} assuming a flat noise spectral density for the detector (in practise we compute the mismatch integral using the \texttt{WaveformMatch} function from the SimulationTools Mathematica package \cite{SimulationTools}). For our sample inspiral with $(p_0,e_0)=(11,0.18)$ we find that the waveform mismatch is always less than $5\times10^{-4}$ for mass ratios in the range $10^{-6} \le \eta \le 10^{-4}$ over durations of 2 months to 2 years -- see Table \ref{table:waveform_mismatch}. In Fig.~\ref{fig:insp_and_waveform} we show the full self-force and NIT inspiral trajectories, waveforms and waveform mismatch for an EMRI with $M=10^{6} M_\odot$ and $\eta=10^{-5}$.

\begin{table}
	\begin{tabular}{ | p{2cm} | p{2cm} | p{2cm} | p{2cm} |}
		\hline
		$\mr$			& 2 months				&	6 months			& 	2 years				\\
		\hline
		$10^{-4}$		& $3.7\times10^{-6}$	&	-					&	-					\\
		$10^{-5}$		& $1.2\times10^{-7}$	& $4.2\times10^{-6}$	& $1.9\times10^{-5}$	\\
		$10^{-6}$		& $2.1\times10^{-9}$	& $5.0\times10^{-7}$	& $4.6\times10^{-4}$	\\
		\hline
	\end{tabular}
	\caption{Waveform mismatch between the self-forced and NIT inspirals for an inspiral with initial parameters $(p_0,e_0) = (11,0.18)$ and $M=10^6 M_\odot$. No data is shown for  the 6 months and 2 years columns for $\eta=10^{-4}$ as this inspiral plunges after $\sim4$ months. The dominant source of error in these results comes from interpolating the inspiral trajectory when computing the waveform. The mismatch can be further reduced, at the expense of computation time, by more densely sampling the NIT inspiral trajectory or using a higher-order interpolation method.}\label{table:waveform_mismatch}
\end{table}

Having demonstrated that NIT inspirals and waveforms faithfully approximate the full self-force results, we now show the rapid speed at which NIT inspirals can be computed. In order to make a fair and detailed comparison between the two methods (and other methods for computing EMRI waveforms) it is worth considering the individual steps in the calculation and their computational time. The three steps are (i) compute the phase space trajectory, (ii) compute the physical trajectory and (iii) compute the waveform. Let us examine each one in turn.

The full self-force equations of motion, \eqref{eq:Seom}, depend upon the orbital phase and so the numerical integrator must take many small steps in order to resolve oscillations on the orbital timescale. Consequently, computing the phase space trajectory $\{p,e,\xi\}$ using the full self-force method takes tens of seconds to hours depending on the initial conditions and the mass-ratio (smaller mass-ratio binaries evolve more slowly and so accumulate more orbits before plunge). As the NIT equations of motion, \eqref{eq:Seomavg}, do not depend on the orbital phase they can be numerically integrated in milliseconds which, depending on the mass ratio, is 2-5 orders of magnitude faster than the full self-force method. In Table \ref{table:speed-up} we give the computation time of the phase space inspirals and the speed up between the full self-force and NIT methods. The millisecond computation time of the NIT model is comparable to kludge methods, but with the benefit of including self-force corrections.

\begin{table}
\begin{tabular}{ | p{1cm} | p{2.2cm} | p{2.2cm} | p{3cm} |}
 \hline
                  & \multicolumn{2}{|c|}{Computation time}	&		 \\
 \hline
 $\mr$ 			  & Full inspiral & NIT inspiral & Inspiral speed up \\
 \hline
 $10^{-3}$        & 6.2s          & 0.008s       & $\sim700$		 \\
 $10^{-4}$        & 43s           & 0.008s  	 & $\sim5,000$		 \\
 $10^{-5}$        & 5m40s		  & 0.008s		 & $\sim40,000$		 \\
 $10^{-6}$        & 42m20s	      & 0.008s		 & $\sim300,000$	 \\
 \hline
\end{tabular}
\caption{Comparison of the phase space trajectory computation time between the full self-force and NIT methods for a variety of mass ratios. All the inspirals start with initial parameters $(p_0,e_0) = (11,0.18)$, or their NIT'd equivalent, and continue to plunge. The time to compute the full inspiral depends on the mass ratio and takes seconds to hours, owing to the need to resolve oscillations on the orbital timescale. By constrast, a NIT inspiral takes milliseconds to compute for any mass ratio. This results in a speed-up of two to five orders of magnitude, depending on the mass ratio.}\label{table:speed-up}
\end{table}

In the full self-force method the computation of the physical trajectory is normally performed simultaneously with solving for the phase space trajectory as the equations for $\{p,e,\xi,t,\phi\}$ form a hierarchically coupled set of equations ($r$ is trivially computed using Eq.~\eqref{eq:Darwin_r} and without loss of generality $\theta=\pi/2$). The addition of the $\{t,\phi\}$ equations adds little to the computation time as their righthand side is cheap to evaluate. Consequently, the second column of Table \ref{table:speed-up} is also indicative of the time to compute the phase space and physical trajectory simultaneously. Computing the physical trajectory using the NIT method is a two-step process. First $\{\tilde{t},\tilde{\phi}\}$ are solved for simultaneously with the phase space variables $\{\tilde{p},\tilde{e},\tilde{\xi}\}$. As with the full self-force method, this adds little computation time and the tilded variables are computed in milliseconds. To compute the physical trajectory, accurate to $\bigO(\eta)$, we need to add the oscillatory $\bigO(\eta^0)$ terms, $Z^{(0)}_{t/\phi}$. These are given analytically in terms of elliptic integrals in Eqs.~\eqref{eq:Z_t0} and \eqref{eq:Z_phi0} and are quick to evaluate. The total time required to compute the physical trajectory in the NIT prescription strongly depends on the sampling rate and duration of the desired waveform. For example, for a 2 month duration equally sampled at 5 second intervals (this equates to $\sim10^6$ samples) computing the physical trajectory takes $\sim0.2$ seconds. For kludge models the time to compute the physical inspiral depends on the model being used. Using the EMRI Kludge Suite \cite{EMRI_Kludge_Suite} implementation of the various kludges, the Analytic Kludge \cite{Barack:2003fp} and Analytic Augmented Kludge \cite{Chua:2017ujo} take around $\sim0.2$ seconds for the same duration and sample rate. The Numerical Kludge takes $\sim4$ seconds, which is longer as it directly solves the $t$ and $\phi$ equations of motion \eqref{eq:Seom_t} and \eqref{eq:Seom_phi}.

Similar to the computation of the physical trajectory, the waveform computation time depends upon the duration and sample rate. Use the semi-relativistic approximation briefly described in Sec.~\ref{sec:waveform_generation} with the same 2 month duration and 5 second sample rate we find the waveform takes $\sim1$ second to compute. This time is the same for both the full self-force and NIT inspiral and is also comparable with kludge methods.

\section{Discussion}\label{sec:conclusion}
In this paper we leveraged the existing machinery of near-identity transformations to obtain equations of motion for small mass-ratio binary systems that are independent of the orbital timescale degrees of freedom. The result is a system of equations that can be evolved at speeds similar to previously developed kludge models, while (in principle) accounting for all physics coming from a systematic expansion of the dynamics in the small mass-ratio --- e.g., gravitational self-forces or corrections due to secondary spin or higher multipoles. As a proof of principle we implemented these equations using the self-forced evolution model of \cite{Warburton:2011fk}. The results show a speed-up of the phase space evolution of 2-5 orders of magnitude compared to evolution of the `full' self-force dynamics, while the phase difference between the two models stays $\bigO(\eta)$. Comparing two-year duration waveforms produced from phase space evolutions from both models, we find that mismatches stay $\lesssim 10^{-4}$.

We must however stress that the implemented model should be viewed as a proof of concept. It does not provide an evolution model that is faithful up to $\bigO(\eta)$ errors in the phases. The main issue is that the model (like that of \cite{Warburton:2011fk} on which it is based) is missing the second-order forcing terms in the osculating geodesics evolution equations. Contributions to these functions include the second order self-force and the post-geodesic corrections to the first order self-force, neither of which have currently been calculated. Calculation of the second order self-force is currently a topic of significant effort \cite{Pound:2012nt,Gralla:2012db,Detweiler:2011tt,Pound:2012dk,Pound:2014xva,Pound:2014koa,Wardell:2015ada,Pound:2015wva,Miller:2016hjv,Pound:2017psq}. Preliminary investigations of the post-geodesic corrections to the first order self-force (using scalar toy models) suggest that the contribution may be negligibly small \cite{Warburton_Capra20}. If not, we note that we are only interested in the corrections to the orbit averaged `fluxes'. In principle, one should be able to calculate this by comparing the time domain fluxes of adiabatic inspirals with the geodesic equivalent fluxes. Once calculations of these contributions become available (which will come at significant one-time offline computational cost), it will be trivial to include them in the NIT averaged inspiral model. Since this model already contains contributions to the second order averaged forcing functions from the oscillatory part of the first order self-force, we expect no additional online computation cost to include these effects in the inspiral calculation.

Another point to note is that our evolution model is applicable only during the inspiral phase of the binary evolution. It will breakdown as the last stable orbit is approached and adiabaticity is lost. This is not an issue for EMRIs detectable by LISA as the plunge, merger, and ringdown phases account for a very small fraction of the SNR compared to the inspiral. For binaries with more comparable mass components the loss of adiabaticity near the last stable orbit will limit the applicability of our approach though how far our model can be pushed towards comparable mass binaries remains to be quantified.

We also note that the implementation here is based on the dataset from \cite{Warburton:2011fk}, which covers only a part of the expected EMRI parameter space for non-spinning binaries. Self-force data for most of the non-spinning parameter space was published in \cite{Osburn:2015duj}. The choice for using the older dataset of \cite{Warburton:2011fk} was motivated by the fact that it included global analytical fits to the data. Consequently, this dataset lent it self well to calculating the phase space derivatives needed for some of the NIT averaged forcing functions. With some care it should also be possible to obtain the phase space derivatives numerically from the data of \cite{Osburn:2015duj}. An alternative approach would be to calculate the phase space derivatives directly when the self-force is computed on a geodesic. This would involve calculating the spacetime derivatives of the self-force and some straight-forward algebra. Both are well within the technological capabilities of the state-of-the-art self-force calculations. This approach may be particularly appealing for filling the EMRI parameter space for spinning binaries. That task will be computationally much more expensive than the non-spinning case, both due to the higher dimensionality of the parameter space and the higher computational costs for calculating the self-force on generic orbits \cite{vandeMeent:2017bcc}.

Although our proof of principle was applied to non-spinning binaries, the averaging NIT can easily accommodate the addition of spin, both to the primary and secondary. When adding spin the equations of motion will involve more than one intrinsic phase. The general derivation of Sec.~\ref{sec:general_NIT} has shown that generically this does not pose any issues. The exception is when the frequencies of the different phases form resonant ratios, in which case the NIT breaks down. Unfortunately, such resonances will appear generically in EMRI evolutions \cite{Flanagan:2010cd,vandeMeent:2013sza,Berry:2016bit,Flanagan:2012kg,Ruangsri:2013hra,Brink:2015roa}. Resonance will therefore need to be dealt with separately. The simplest approach would be to simply switch back to the full evolution equations just before hitting the resonance, evolving through the resonance, and switching back to the NIT averaged equations. This will undoubtedly work, but will come at a significant computational cost. We can already do a lot better by rather than switching back to the full equations of motion, using a NIT to eliminate all non-resonant oscillating terms as in \cite{vandeMeent:2013sza}. By definition the resonant terms will only vary slowly in the vicinity of the resonance limiting the computational cost. However, the best solution would be obtained if one could implement the effects of the resonance as an instantaneous jump on the orbital parameters. The results of \cite{Flanagan:2010cd} and \cite{vandeMeent:2013sza} suggest that this may be possible. In any case, however, the inclusion of resonances in fast-evolution models will require further consideration in future work.

Finally, we note that in this work we employed a simple model to produce a gravitational waveform from the inspiral dynamics. For our current purposes this was sufficient, as we were using the same waveform generation scheme for both evolutionary models that were compared. However, for application in EMRI data analysis a more realistic, and fast to compute, waveform model will be needed. One approach would be to utilize the two-timescale expansion of the waveform at infinity. At leading order this will be given by a function $h(\vec{P}(t),\vec{q}(t),\vec{S}(t))$ \cite{Hinderer:2008dm}, which can be rewritten as function $\tilde{h}(\vec{\tilde{P}}(t),\vec{\tilde{q}}(t),\vec{\tilde{S}}(t))$. It is worth investigating whether an efficient numerical surrogate for this function can be build from the waveforms generated by particles on geodesic orbits. This will be pursued in future work.

\begin{acknowledgments}
MvdM was supported by European Union's Horizon 2020 research and innovation programme under grant agreement No.~705229. NW gratefully acknowledges support from a Royal Society - Science Foundation Ireland University Research Fellowship. We thank Ian Hinder and Barry Wardell for the SimulationTools analysis package.
\end{acknowledgments}

\section*{References}
\raggedright
\bibliography{NITevolve,journalshortnames,commongsf,meent}

\appendix

\section{Schwarzschild forcing functions}\label{app:pars}
The gravitational self-force (GSF) is defined as the correction to the geodesic equation for the trajectory of an object due to the object's own influence on the gravitational field. Formally, this has a functional dependence on the past trajectory of the object. However, assuming no gravitational waves coming in from past null infinity and a fixed gauge, there should be a unique trajectory going through point each event $x^\alpha$ in the background spacetime for each four-momentum $p_\beta$ at that event. By taking this trajectory as the past for any point in $(x^\alpha, p_\beta)$ in the (test particle) phase space, the  GSF can be written as a (local) function of $(x^\alpha, p_\beta)$ giving a closed and local equation of motion,
\begin{equation}\label{eq:gdGSF}
\CD{u}p_\mu =\mr^2 \gsf_\mu(x^\alpha,p_\beta) ,
\end{equation}
where we have extracted a factor of the mass-ratio squared such that we expect  $\gsf_\mu = \bigO(\eta^0)$. At leading order in $\eta$, one expects $\gsf_\mu(x^\alpha,p_\beta)$ to coincide with the GSF generated by a particle whose past trajectory is a geodesic through $(x^\alpha,p_\beta)$. Using the osculating geodesic formalism \eqref{eq:gdGSF} can be rewritten as a set of first order equations for the geodesic  elements $(p,e,\xi,t,\phi)$ \cite{Pound:2007th},
\begin{subequations}
\begin{align}
\d{\xi}{\chi} &= 1 + \eta f_\xi(p,e,\xi), \\
\d{p}{\chi} &= \eta \mathcal{F}_p(p,e,\xi),\\
\d{e}{\chi} &= \eta \mathcal{F}_e(p,e,\xi),\\
\d{t}{\chi} &= \omega_t(p,e,\xi), \\
\d{\phi}{\chi} &= \omega_\phi(p,e,\xi),
\end{align}
\end{subequations}
with

\begin{align}
\mathcal{F}_p &= \frac{2p^{3}M(p-3-e^2)}{b_{+}^2 b_{-}^2(1+ e \cos\xi)^2}
 \hh{
 	\frac{p^{1/2}M b_{\xi}(p-3-e^2\cos^2\xi)}{(1+ e \cos\xi)^2} \gsf^\phi
	- e\sin\xi \gsf^r
 },
\\
\mathcal{F}_e &= 
\frac{-2p^{5/2}M^2 (p-3-e^2)}{b_{+}^2 b_{-}^2(1+ e \cos\xi)^2}\Bh{
\frac{(p-6-2e^2)\sin\xi}{M p^{1/2}} \gsf^r
\\ \notag &\qquad
-\frac{(p-6-2e^2)\cos\xi\Bh{b_{\xi}^2e\cos\xi+2(p-3)}
+e(p^2-10p+12+4 e^2)
}{b_{\xi}(1+ e \cos\xi)^2} \gsf^\phi
},
\\
f_\xi &=
\frac{-p^{5/2}M^2 (p-3-e^2)}{e b_{+}^2 b_{-}^2 (1+ e \cos\xi)^2}
\Bh{
\frac{\bh{(p-6)\cos\xi+2e}}{p^{1/2}M} \gsf^r
\\ \notag
&\qquad-\frac{
\sin\xi\Bh{(p-6)\bh{b_{\xi}^2 e\cos\xi+2(p-3)}-4e^3\cos\xi}
}{
b_{\xi}(1+ e \cos\xi)^2
}\gsf^\phi
},
\\
 \omega_t &= \frac{a_{+}a_{-} p^2}{a_\xi^2 b_\xi (1+ e \cos\xi)^2},\\
 \omega_\phi &=  \frac{\sqrt{p}}{b_\xi},
\end{align}
and where we introduced the following shorthand,
\begin{align}
a_{+} &:= \sqrt{p-2+2e},\\
a_{-} &:= \sqrt{p-2-2e},\\
a_{\xi} &:= \sqrt{p-2-2e\cos\xi},\\
b_{+} &:= \sqrt{p-6+2e},\\
b_{-} &:= \sqrt{p-6-2e},\\
b_{\xi} &:= \sqrt{p-6-2e\cos\xi}.
\end{align}

\section{An Explicit NIT in Schwarzschild spacetime}\label{app:NIT}
The (near-identity) transform needed to reach the averaged form of the equations of motion \eqref{eq:Seomavg} is given by,
\begin{subequations}\label{eq:Snit}
\begin{alignat}{4}
\tilde{\xi} &= \xi &&+\eta X^{(1)}(p,e,\xi) &&+\eta^2  X^{(2)}(p,e,\xi) +\bigO(\eta^3),\\
\tilde{p} &=p &&+\eta Y_p^{(1)}(p,e,\xi) &&+\eta^2  Y_p^{(2)}(p,e,\xi) +\bigO(\eta^3),\\
\tilde{e} &= e &&+\eta Y_e^{(1)}(p,e,\xi) &&+\eta^2  Y_e^{(2)}(p,e,\xi) +\bigO(\eta^3),\\
\tilde{t} &= t + Z_t^{(0)}(p,e,\xi) &&+\eta Z_t^{(1)}(p,e,\xi) &&+\bigO(\eta^2),\\
\tilde{\phi} &= \phi + Z_\phi^{(0)}(p,e,\xi) &&+\eta Z_\phi^{(1)}(p,e,\xi) &&+\bigO(\eta^2).
\end{alignat}
\end{subequations}
The zeroth order  functions $Z_{t/\phi}^{(0)}$ are defined by the equations,
\begin{align}
\pd{Z_t^{(0)}}{\xi}(p,e,\xi) &= -\breve{\omega}_t(p,e,\xi), \\
\pd{Z_\phi^{(0)}}{\xi}(p,e,\xi) &= -\breve{\omega}_\phi(p,e,\xi).
\end{align}
This can be solved analytically  in terms of elliptic functions \cite{Darwin:1959,Darwin:1961},
\begin{align}
Z_t^{(0)}(p,e,\xi) &= 														\label{eq:Z_t0}
\frac{p a_{+}a_{-}}{(1-e^2)b_{+}}\ellF{\frac{\xi-\pi}{2}}{k_r}
-\frac{pa_{+}a_{-}b_{+}}{(1-e^2)(p-4)}\ellE{\frac{\xi-\pi}{2}}{k_r}
\\ \notag
&\qquad  
- 2\hh{\frac{a_{+}^2a_{-}^2}{(1-e^2)(p-4)}+3}\frac{a_{+}a_{-}}{(1-e)b_{+}}\ellPi{-\frac{2e}{1-e}}{\frac{\xi-\pi}{2}}{k_r}
\\ \notag
&\qquad 
- 8\frac{a_{-}}{a_{+}b_{+}}\ellPi{\frac{4e}{a_{+}^2}}{\frac{\xi-\pi}{2}}{k_r}
\\ \notag
&\qquad 
+\frac{T_r(p,e)}{2\pi}(\xi-\pi)+ \frac{e p a_{+}a_{-}b_\xi}{(1-e^2)(p-4)(1+e\cos\xi)} \sin\xi		
\\
Z_\phi^{(0)}(p,e,\xi) &=\frac{\Phi_r(p,e)}{2\pi}(\xi-\pi) -2\frac{\sqrt{p}}{b_{+}}\ellF{\frac{\xi-\pi}{2}}{k_r},		\label{eq:Z_phi0}
\end{align}
where $\ellF{\varphi}{k}$, $\ellE{\varphi}{k}$, and $\ellPi{h}{\varphi}{k}$ are elliptic functions of first, second, and third kind (following the conventions for the arguments used in \emph{Mathematica}), and we introduced the short-hand
\begin{align}
k_r &:=  \frac{4e}{p-6+2e}.
\end{align}
Note that although the expressions for $Z_{t/\phi}^{(0)}$ contain explicit linear terms $\xi$, these are canceled by  secular contributions from the elliptic functions, and as a whole the $Z_{t/\phi}^{(0)}$ are purely oscillatory functions of $\xi$.

Finally $T_r(p,e)$ and $\Phi_r(p,e)$ are the radial period and the accumulated $\phi$ over one such period or a geodesic with semi-latus rectum $p$ and eccentricity $e$,
\begin{align}
T_r(p,e) &= \label{eq:T_r}
\frac{2p a_{+}a_{-}b}{(1-e^2)(p-4)}\ellEc{k_r} 
-2p \frac{a_{+}a_{-}}{(1-e^2)b}\ellK{k_r}
+\frac{16a_{-}}{a_{+}b}\ellPic{\frac{4e}{p-2+2e}}{k_r}
\\
&\qquad 
- \frac{4\hh{8(1-e^2)+p(1+3e^2-p)}a_{+}a_{-}}{(1-e)(1-e^2)(p-4)b}\ellPic{-\frac{2e}{1-e}}{k_r} \nonumber,
\\
\Phi_r(p,e) &= 4\frac{\sqrt{p}}{b}\ellK{k_r} \label{eq:Phi_r}.
\end{align}
The $\bigO(\eta)$ terms in the transformation are given by
\begin{align}
X^{(1)} &= \avg{X^{(1)}}-\int \breve{f_\xi} \id\xi,\\
Y_p^{(1)} &= -\int \breve{F_p} \id\xi,\\
Y_e^{(1)} &= -\int \breve{F_e} \id\xi,\\
Z_t^{(1)}
 &= 
 \int\Bh{
  \osc{\breve{\omega}_t f_\xi}
- \osc{\pd{\breve{Z_t}^{(0)}}{p} F_p}
- \osc{\pd{\breve{Z_t}^{(0)}}{e} F_e}
\\ \notag&\qquad\qquad
- \frac{1}{2\pi}\pd{T_r}{p}\int \breve{F_p} \id\xi
- \frac{1}{2\pi}\pd{T_r}{e}
}\id\xi,
 \\
Z_\phi^{(1)}
 &=
 \int\Bh{
 \osc{\breve{\omega}_\phi  f_\xi}
-\osc{\pd{\breve{V}^{(0)}}{p} F_p}
-\osc{\pd{\breve{V}^{(0)}}{e} F_e}
\\ \notag&\qquad\qquad
-\frac{1}{2\pi}\pd{\Phi_r}{p}\int \breve{F_p} \id\xi
-\frac{1}{2\pi}\pd{\Phi_r}{e}\int \breve{F_e} \id\xi
}\id\xi,
\end{align}
where $ \avg{X^{(1)}}$ satisfies the first-order PDE (no explicit solution is needed anywhere),
\begin{align}
 \pd{\avg{X^{(1)}}}{p} \avg{F_p}
 +\pd{\avg{X^{(1)}}}{e} \avg{F_e}
 =
 \avg{  \breve{F}_p\int \pd{\breve{f_\xi}}{p} \id\xi}
 +\avg{ \breve{F}_e\int \pd{\breve{f_\xi}}{e} \id\xi }
 +\avg{\breve{f}_\xi \breve{f}_\xi}
\end{align}
and the primitive $\int \cdot \id\xi$ of a purely oscillatory is chosen to be purely oscillatory.  That is, given a Fourier decomposition 
\begin{align}
\breve{A}	&=\sum_{\kappa\neq0} A_{\kappa}e^{\ii \kappa\xi},
\end{align}
its primitive is given by
\begin{align}
\int\breve{A}\id\xi	&=\sum_{\kappa\neq0} \frac{A_{\kappa}}{\ii\kappa}e^{\ii \kappa\xi}.
\end{align}
Finally, the second order terms in the transformation are given by
\begin{align}
X^{(2)}
 &=
 \int\Bh{
 \osc{\breve{f}_\xi f_\xi}
 +\osc{F_p \int \pd{ \breve{f_\xi} }{p} \id\xi }
 +\osc{F_e \int \pd{ \breve{f_\xi} }{e} \id\xi }
 -\breve{F}_p \pd{\avg{X^{(1}}}{p} 
 -\breve{F}_e \pd{\avg{X^{(1}}}{e}
 \\ \notag &\qquad\qquad
 -\pd{\avg{f_\xi}}{p} \int \breve{F_p} \id\xi
 -\pd{\avg{f_\xi}}{e} \int \breve{F_e} \id\xi
 }\id\xi,
 \\
Y_p^{(2)}
  &=
   \int\Bh{
\osc{\breve{F}_p f_\xi}
 +\osc{F_p \int \pd{\breve{F_p}}{p} \id\xi }
 +\osc{F_e \int \pd{\breve{F_p}}{e} \id\xi }
  \\ \notag &\qquad\qquad
 -\pd{\avg{F_p}}{p} \int \breve{F_p} \id\xi
 -\pd{\avg{F_p}}{e} \int \breve{F_e} \id\xi
  }\id\xi,
 \\
Y_e^{(2)}
  &=
 \int\Bh{
 \osc{\breve{F}_e f_\xi}
 +\osc{F_p\int  \pd{\breve{F_e}}{p} \id\xi }
 +\osc{F_e\int  \pd{\breve{F_e}}{e} \id\xi }
  \\ \notag &\qquad\qquad
 -\pd{\avg{F_e}}{p}\int \breve{F_p} \id\xi
 -\pd{\avg{F_e}}{e} \int \breve{F_e} \id\xi
  }\id\xi.
\end{align}

To conclude this appendix we give explicit expressions for the averaged forcing functions in terms of the Fourier coefficients of the original forcing functions. Expanding the original forcing terms using Eq.~\eqref{eq:Fourier_decomp}, the averaged forcing functions in \eqref{eq:tilde_Fs} are given by:

\begin{subequations}\label{eq:tilde_Fsexplicit}
\begin{align}
\tilde{f}_\xi^{(1)} &= \avg{f_\xi} = f_{\xi,0} ,
\\
\tilde{\mathcal{F}}_p^{(1)} &=  \avg{ \mathcal{F}_p}= \mathcal{F}_{p,0},
\\
\tilde{\mathcal{F}}_e^{(1)} &= \avg{ \mathcal{F}_e}= \mathcal{F}_{e,0},
\\
\tilde{f}_t^{(1)} &= 
 \avg{\pd{\breve{Z}_t^{(0)}}{p} \breve{F}_p}
+ \avg{\pd{\breve{Z}_t^{(0)}}{e} \breve{F}_e} 
= 
\sum_{\kappa\neq0}\Bh{\pd{Z_{t,-\kappa}^{(0)}}{p} F_{p,\kappa}+\pd{Z_{t,-\kappa}^{(0)}}{e} F_{e,\kappa} },
\\
\tilde{f}_\phi^{(1)} &= 
 \avg{\pd{\breve{Z}_\phi^{(0)}}{p} \breve{F}_p}
+ \avg{\pd{\breve{Z}_\phi^{(0)}}{e} \breve{F}_e}
= 
\sum_{\kappa\neq0}\Bh{\pd{Z_{\phi,-\kappa}^{(0)}}{p} F_{p,\kappa}+\pd{Z_{\phi,-\kappa}^{(0)}}{e} F_{e,\kappa} },
\\
\tilde{\mathcal{F}}_p^{(2)} &=
 -\avg{\breve{\mathcal{F}}_p \int \pd{ \breve{\mathcal{F}_p}}{p}\id\xi }
 -\avg{\breve{\mathcal{F}}_e \int \pd{\breve{\mathcal{F}_p}}{e}\id\xi }
 -\avg{\breve{\mathcal{F}}_p \breve{f}_\xi}
\\ \notag
 &= -\sum_{\kappa\neq 0}
 \Bh{ 
 \frac{\ii}{\kappa}\mathcal{F}_{p,\kappa} \pd{ \mathcal{F}_{p,-\kappa}}{p} 
 +\frac{\ii}{\kappa}\mathcal{F}_{e,\kappa} \pd{ \mathcal{F}_{p,-\kappa}}{e} 
 +\mathcal{F}_{p,\kappa}f_{\xi,-\kappa}
 },
\\
\tilde{\mathcal{F}}_e^{(2)} &=
 -\avg{\breve{\mathcal{F}}_p\int \pd{\breve{\mathcal{F}_e}}{p} \id\xi }
 -\avg{\breve{\mathcal{F}}_e\int \pd{\breve{\mathcal{F}_e}}{e} \id\xi }
 -\avg{\breve{\mathcal{F}}_e \breve{f}_\xi}
 \\ \notag
 &= -\sum_{\kappa\neq 0}
 \Bh{ 
 \frac{\ii}{\kappa}\mathcal{F}_{p,\kappa} \pd{ \mathcal{F}_{e,-\kappa}}{p} 
 +\frac{\ii}{\kappa}\mathcal{F}_{e,\kappa} \pd{ \mathcal{F}_{e,-\kappa}}{e} 
 +\mathcal{F}_{e,\kappa}f_{\xi,-\kappa}
 }.
\end{align}
\end{subequations}

\end{document}